\documentclass[lettersize,journal]{IEEEtran}
\usepackage{amsmath,amsfonts,amssymb}
\usepackage{physics}
\usepackage{algorithmic}
\usepackage{algorithm}
\usepackage{array}
\usepackage{textcomp}
\usepackage{stfloats}
\usepackage{url}
\usepackage{verbatim}
\usepackage{graphicx}
\usepackage{cite}
\usepackage{csquotes}
\usepackage{booktabs}
\usepackage[nolist, nohyperlinks]{acronym}
\usepackage{xcolor}
\usepackage{xfrac}
\usepackage{caption}
\usepackage{subcaption}
\usepackage{microtype}

\usepackage[subtle]{savetrees}

\usepackage[pagebackref,breaklinks,colorlinks]{hyperref}
\usepackage{orcidlink}

\usepackage[capitalize]{cleveref}

\DeclareMathOperator*{\argmin}{arg\,min}
\newcommand{\D}{\mathrm{d}}
\newcommand{\w}{\mathbf{w}}
\newcommand{\f}{\mathbf{f}}
\newcommand{\x}{\mathbf{x}}
\newcommand{\y}{\mathbf{y}}
\newcommand{\z}{\mathbf{z}}
\newcommand{\bmu}{\boldsymbol{\mu}}
\newcommand{\submin}{_{\text{min}}}
\newcommand{\submax}{_{\text{max}}}
\newcommand{\Reals}{\mathbb{R}}

\newcommand{\celeba}[0]{CelebA-HQ256}

\newcommand{\B}[1]{\textbf{#1}}
\definecolor{darkred}{rgb}{0.64, 0.0, 0.0}

\newcommand{\psnrb}[0]{PSNR\nobreakdash-B}

\begin{acronym}
\acro{sgm}[SGM]{score-based generative model}
\acro{snr}[SNR]{signal-to-noise ratio}
\acro{gan}[GAN]{generative adversarial network}
\acro{vae}[VAE]{variational autoencoder}
\acro{ddpm}[DDPM]{denoising diffusion probabilistic model}
\acro{stft}[STFT]{short-time Fourier transform}
\acro{istft}[iSTFT]{inverse short-time Fourier transform}
\acro{sde}[SDE]{stochastic differential equation}
\acro{ode}[ODE]{ordinary differential equation}
\acro{ou}[OU]{Ornstein-Uhlenbeck}
\acro{ve}[VE]{Variance Exploding}
\acro{ouve}[OUVE]{Ornstein-Uhlenbeck Variance Exploding}
\acro{tsdve}[TSDVE]{t-squared Decay Variance Exploding}
\acro{vp}[VP]{Variance Preserving}
\acro{dnn}[DNN]{deep neural network}
\acro{pesq}[PESQ]{Perceptual Evaluation of Speech Quality}
\acro{se}[SE]{speech enhancement}
\acro{tf}[T-F]{time-frequency}
\acro{elbo}[ELBO]{evidence lower bound}
\acro{WPE}{weighted prediction error}
\acro{PSD}{power spectral density}
\acro{RIR}{room impulse response}
\acro{SNR}{signal-to-noise ratio}
\acro{LSTM}{long short-term memory}
\acro{POLQA}{Perceptual Objective Listening Quality Analysis}
\acro{SDR}{signal-to-distortion ratio}
\acro{ESTOI}{Extended Short-Term Objective Intelligibility}
\acro{ELR}{early-to-late reverberation ratio}
\acro{TCN}{temporal convolutional network}
\acro{DRR}{direct-to-reverberant ratio}
\acro{nfe}[NFE]{number of function evaluations}
\acro{rtf}[RTF]{real-time factor}
\acro{dsm}[DSM]{Denoising Score Matching}
\acro{qgac}[QGAC]{Quantization Guided JPEG Artifact Correction}
\acro{fbcnn}[FBCNN]{Flexible Blind Convolutional Neural Network}
\acro{fid}[FID]{Fréchet Inception Distance}
\acro{kid}[KID]{Kernel Inception Distance}
\acro{lpips}[LPIPS]{Learned Perceptual Image Patch Similarity}
\acro{ssim}[SSIM]{Structural Similarity}
\acro{qf}[QF]{Quality Factor}
\acro{ddrm}[DDRM]{Denoising Diffusion Restoration Model}
\acro{psnr}[PSNR]{Peak Signal-to-Noise Ratio}
\acro{bef}[BEF]{Blocking Effect Factor}
\end{acronym}

\begin{document}

\title{DriftRec: Adapting diffusion models\\to blind JPEG restoration}

\author{Simon Welker$^{1,2}$ \orcidlink{0000-0002-6349-8462}, Henry N. Chapman$^{2,3,4}$ \orcidlink{0000-0002-4655-1743}, Timo Gerkmann$^1$ \orcidlink{0000-0002-8678-4699}\\
{\small
    $^1$Signal Processing (SP), Department of Informatics, Universität Hamburg, Germany\\
    $^2$Center for Free-Electron Laser Science CFEL, Deutsches Elektronen-Synchrotron DESY, Notkestr. 85, 22607 Hamburg, Germany\\
    $^3$Centre for Ultrafast Imaging, Universität Hamburg, Germany\\
    $^4$Department of Physics, Universität Hamburg, Germany\\
}%
}%

\maketitle

\begin{abstract}
In this work, we utilize the high-fidelity generation abilities of diffusion models to solve blind JPEG restoration at high compression levels. We propose an elegant modification of the forward stochastic differential equation of diffusion models to adapt them to this restoration task and name our method \emph{DriftRec}. Comparing DriftRec against an $L_2$ regression baseline with the same network architecture and state-of-the-art techniques for JPEG restoration, we show that our approach can escape the tendency of other methods to generate blurry images, and recovers the distribution of clean images significantly more faithfully. For this, only a dataset of clean/corrupted image pairs and no knowledge about the corruption operation is required, enabling wider applicability to other restoration tasks. In contrast to other conditional and unconditional diffusion models, we utilize the idea that the distributions of clean and corrupted images are much closer to each other than each is to the usual Gaussian prior of the reverse process in diffusion models. Our approach therefore requires only low levels of added noise and needs comparatively few sampling steps even without further optimizations. We show that DriftRec naturally generalizes to realistic and difficult scenarios such as unaligned double JPEG compression and blind restoration of JPEGs found online, without having encountered such examples during training.
\end{abstract}


\section{Introduction}

Diffusion models have taken the world of machine learning by storm due to their unprecedented ability to generate high-fidelity images~\cite{dhariwal2021diffusion,ramesh2022hierarchical,rombach2022high} and audio~\cite{kong2021diffwave,richter2022journal,huang2022fastdiff}, their relatively easy training, and their great flexibility to effectively condition on a variety of user inputs~\cite{rombach2022high,sheffer2022hear,ramesh2022hierarchical}. Previous works on diffusion models have largely concentrated on unconditional and conditional image generation. Recently, Bansal et al. \cite{bansal2022cold} and Daras et al. \cite{daras2023soft} have proposed to extend and improve the forward and reverse processes of diffusion models by employing known deterministic corruptions. However, they have only evaluated their ideas for unconditional generation tasks, rather than for faithfully inverting corruptions to restore plausible original images. Here, we pursue a different route with our method \emph{DriftRec}, and explicitly modify and train task-adapted diffusion models to restore natural-looking images from JPEG-compressed ones.

In contrast to other works~\cite{bansal2022cold,daras2023soft,kawar2022denoising}, which make at least one of the following assumptions, DriftRec does not require the underlying corruption operator to be known, linear, nor from a continuous family of operators. Instead, it requires only a dataset of (clean image, corrupted image) pairs, as in classic supervised training. DriftRec utilizes an alternative way of combining a deterministic corruption with the usual Gaussian noise used in diffusion models, within the formalism based on \acp{sde} introduced by Song et al.~\cite{song2021sde}. In this work, we focus on blind JPEG restoration, based upon the observation that JPEG compression is a corruption with a nonlinear and nondifferentiable forward operator that may be unknown to the restoration method at inference time in many practical settings. We propose a modification of the forward \ac{sde} of diffusion models to adapt them to image restoration, building upon our previous work in the speech processing literature \cite{welker2022speech,richter2022journal} and extending our previous workshop contribution~\cite{welker2022blind} with additional datasets, other forward \acp{sde}, and substantially more experiments. Our code is available at \url{https://github.com/sp-uhh/driftrec}.

\subsection{Related work}\label{sec:related-work}
For image restoration based on pretrained unconditional diffusion models, the formalism of \acp{ddrm}~\cite{kawar2022denoising} has been proposed. \acp{ddrm} were originally only designed for linear inverse problems, but have very recently also been adapted to JPEG restoration by carefully considering JPEG encoding and decoding operators~\cite{kawar2022jpeg}. The essential proposed advantage of these methods is the ability to train only one unconditional diffusion model and use it for a variety of restoration tasks. While this is advantageous from the standpoint of computational efficiency, these methods require careful task-adapted constructions and rely on intricate knowledge of the corruption operator being available at inference time. In contrast, our approach requires only a paired dataset for supervised training and does not need the corruption operator to be known, neither during training nor during inference.

In~\cite{saharia2022palette}, the authors devise a diffusion model for image-to-image translation called \emph{Palette}, and also briefly evaluate their approach for blind JPEG restoration. In contrast to our work, Palette does not adapt the stochastic processes to the task but simply uses the corrupted images as conditioning information to generate reconstructions, letting the sampling process start from an unstructured Gaussian prior. Their approach uses 10 times as many DNN evaluations as our method, impacting runtime by a similar factor. Palette also uses around 8.5 times as many DNN parameters (552\,M) as our models.

For JPEG restoration specifically, recent methods often utilize knowledge about the corruption operation, such as \aclu{qf} estimates~\cite{jiang2021towards} or the quantization matrices stored in each image~\cite{ehrlich2020quantization,kawar2022jpeg}. Methods that rely on JPEG-internal information being available cannot easily be applied to many images typically encountered online, which have been edited, re-compressed, or saved in other formats before or after JPEG compression. We intentionally design our approach such that it can be applied to complex corruptions that are not known or easily modeled, and to determine the achievable performance in the fully blind setting. Other recent blind approaches include the optimization- and unfolding-based methods \cite{fu2021model} and \cite{fu2021learning} which, in contrast to our method, have however both only been trained and tested for a small selection of \acp{qf}, and FBCNN \cite{jiang2021towards} which, as we will show, degrades in the blind doubly-compressed JPEG scenario and tends to produce blurry outputs for low \acp{qf}.

Luo et al.~\cite{luo2023image} use an approach closely related to ours~\cite{welker2022speech,richter2022journal,welker2022blind,welker2022driftrec} and demonstrate general applicability to a wide array of restoration tasks such as deraining, deblurring, denoising, super-resolution, and inpainting. They also propose a different training objective and different family of forward \acp{sde}. Here, we focus more deeply on a specific task (JPEG restoration) and offer additional insights into possible forward \acp{sde}, generalization to real-world restoration scenarios, sampler choices, sample averaging, and improvements for a downstream classifier. Luo et al.~\cite{luo2023image} note that their family of forward \acp{sde} results in an oversmooth variance of the forward process. We offer a numerical solution to this problem in \cref{sec:numerical-forward-sde} and use it to implement our \emph{CosVE} \ac{sde}.

Liu et al.~\cite{liu2023i2sb} derive a Schrödinger Bridge formalism for image-to-image tasks called \emph{I2SB}, using a linearly rising and decaying noise schedule. I2SB effectively results in a different choice of \ac{sde} than ours (see \cref{sec:sde-diffusion-models}), which enjoys the property of terminating exactly at the corrupted image. The authors show that compared to Palette, the \ac{nfe} can be significantly reduced at little decrease in quality, thereby confirming the advantages of task-adapted diffusion processes in line with our arguments. However, the authors only consider each restoration task briefly and use highly specific, effectively non-blind models (e.g., one DNN per JPEG \ac{qf}). In our Results section, we show that in a fair comparison, our method outperforms I2SB for general JPEG restoration.

Delbracio et al.~\cite{delbracio2023inversion} propose a different, deterministic way of iterative image restoration not based directly on a diffusion formalism, designing a continuous interpolation between each clean and degraded image pair similar to the mean of the processes in this paper, see \cref{eq:mu-ouve} as well as the related works \cite{liu2023i2sb,luo2023image}. They discuss connections to image-to-image diffusion models including our method \cite{welker2022blind}, and interestingly find that injecting stochasticity back into their deterministic system can result in significant quality improvements.

\section{Methods}

\subsection{SDE-based diffusion models}
\label{sec:sde-diffusion-models}

Following \cite{song2021sde}, the forward process of a diffusion model can be interpreted as a dynamical system that follows an \ac{sde}
\begin{equation}\label{eq:forward-sde}
    \D{\x_t} = \f(\x_t, t) \D{t} + g(\x_t, t) \D{\w}
\end{equation}
where in this work, $\x_t \in \Reals^{C \times H \times W}$ is the current image and $\w$ is a standard Wiener process of the same dimensionality as $\x_t$, and the process runs forward from $t=t_\varepsilon$ until the terminal process time $T:=1$ (with $t_\varepsilon \gtrapprox 0$ for reasons of numerical stability \cite{song2021sde}). $f$ is called the \emph{drift coefficient} and is deterministic, while $g$ is called the \emph{diffusion coefficient} and controls the strength of the stochastic Wiener process.
Each image in the training dataset then represents the initial value $\x_0$ of a particular realization of this \ac{sde}. Song et al. \cite{song2021sde} showed that previous diffusion models in the discrete-time domain can be interpreted to follow either the so-called \emph{\ac{ve} \ac{sde}} or the so-called \emph{\ac{vp} \ac{sde}}. Both \acp{sde} have the aim of progressively turning images into Gaussian white noise, thereby turning the intractable image distribution into a tractable prior. To generate images, one then samples an initial value from this prior and numerically solves the corresponding \emph{Reverse \ac{sde}} \cite{anderson1982reverse},
\begin{equation}\label{eq:reverse-sde}
    \D{\x_t} = [-\f(\x_t, t) + g(t)^2 \grad_{\x_t} \log p_t(\x_t, t)] \D{t} + g(\x_t, t) \D{\bar{\w}}
\end{equation}
where $\bar{\w}$ is a Wiener process running in reverse, and we assume an infinitesimal $\D{t}$ with a positive sign. The only unknown term in \eqref{eq:reverse-sde} is the \emph{score} $\grad_{\x_t} \log p_t(\x_t, t)$. A deep neural network called a \emph{score network} $S_\theta(\x_t, t)$ is then trained to estimate this score, given the current process state $\x_t$ with diffusion time $t$. By replacing the score with its learned approximation $S_\theta$, we receive the so-called \emph{plug-in reverse \ac{sde}}. To generate samples, this plug-in reverse \ac{sde} is then solved with numerical \ac{sde} solver schemes~\cite{song2021sde,sarkka2019sde,kloeden2011numerical}.

\subsection{A family of task-adapted linear SDEs}\label{sec:task-adapted-sdes}

Rather than turning the clean image distribution into pure noise, we adapt our forward process to turn the clean image distribution into a noisy version of the corrupted image distribution. This has two purposes:
\begin{enumerate}
    \item Instead of pure noise, this uses the corrupted image (plus tractable noise) as the initial value of the reverse \ac{sde}, thus achieving the task adaptation through the formulation of the process itself, as opposed to only providing the corrupted image as conditioning information.
    \item Since the added Gaussian noise is white, it functions as a continual source of all possible spatial frequencies throughout the reverse process. The trained score model then filters these frequencies appropriately to generate plausible clean images from the target distribution without a loss of high-frequency detail.
\end{enumerate}
Note that the distribution of noisy corrupted images is still tractable for the restoration task, as a sample from it can be drawn by taking a corrupted image and adding Gaussian noise. Some other works also use similar initialization strategies, as a trick to speed up inference~\cite{kawar2022jpeg} or to generate images from rough sketches~\cite{meng2022sdedit}, but without adapting the \ac{sde}. While empirically this seems to work reasonably well, it is formally (and in principle, arbitrarily strongly) mismatched to the actual distributions encountered throughout the reverse process. Our \ac{sde}-based adaptation of the process largely avoids this problem by ensuring that this initialization corresponds to the terminal distribution of the forward process which is also the initial distribution of the reverse process.

\acp{sde} which have a drift that is an affine function of the process state $x_t$ have the advantageous property that they admit a closed-form solution that is fully characterized by the mean and variance of the forward process~\cite[Sec.~5]{sarkka2019sde}. This property enables efficient calculation of samples of the forward process at arbitrary process times $t$, without having to solve the forward \ac{sde} numerically for each training sample, which is necessary for efficient training. We propose the following family of linear forward \acp{sde} to realize our idea:
\begin{equation}\label{eq:forward-drift-sde-family}
    \f(\x_t, \y, t) = \gamma F(t) \cdot (\y - \x_t)\,, \hspace{1em}
    g(t) = \nu \left( \frac{\sigma\submax}{\sigma\submin} \right)^{2t}\,,
\end{equation}
where $\y$ is the corrupted image corresponding to $\x_0$, $\gamma$ is a stiffness hyperparameter controlling how strongly $\x_t$ is pulled towards $\y$, and $F(t)$ -- here a scalar function -- controls the shape of the curve pulling $\x_t$ towards $\y$. $\nu$ is a normalization factor determined to ensure that $\sigma_T \approx \sigma\submax$, where $\sigma_t$ is the closed-form variance of the Gaussian process described by \eqref{eq:forward-drift-sde-family}.

Intuitively, our family of \acp{sde} combines the diffusion $g$ of the \acf{ve} \ac{sde}~\cite{song2019generative,song2021sde} with an added drift term $\f$ that pulls $\x_t$ towards the corrupted image $\y$. We choose the diffusion term of the \ac{ve} \ac{sde} since this \ac{sde} has shown to result in high sample quality~\cite{song2021sde}, but in principle, other diffusion terms could be used here, as done in~\cite{luo2023image}.

\begin{figure*}[t]
    \centering
    \includegraphics[width=\textwidth]{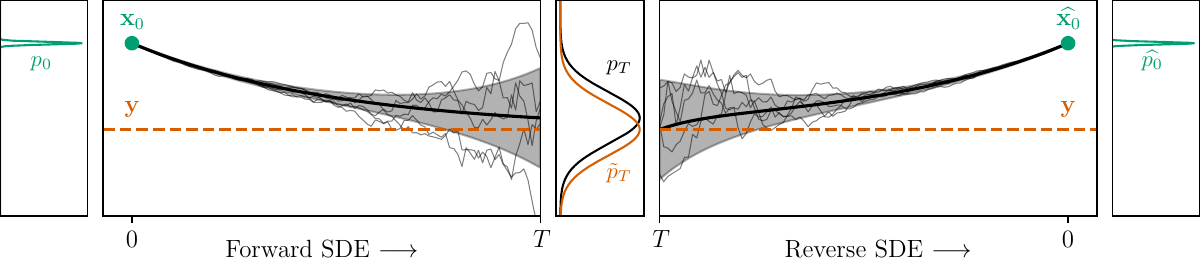}
    \caption{Reproduced from~\cite{richter2022journal}: The forward and reverse process for the \ac{ouve} \ac{sde} illustrated with a single scalar variable. The mean (thick black line) of the forward process exponentially decays from a clean sample $\mathbf x_0$ (blue) towards the corrupted sample $\mathbf y$ (green), and the standard deviation (shaded gray region) increases exponentially. The reverse process then starts from a slightly mismatched distribution $\tilde{p}_T$ which is centered around $\mathbf y$ rather than $\mathbf x_T$ and moves to an estimate $\hat{\x_0}$. Five realizations of both processes are shown as thin black lines.}
    \label{fig:forward-reverse-process}
\end{figure*}

It is interesting to note that the drift term $\f$ of the forward SDE \eqref{eq:forward-sde} shows up with a \emph{negative} sign in the reverse SDE \eqref{eq:reverse-sde}. This term, $-\f$, continually pushes the process state $\x_t$ away from $\y$ during the reverse process. In \cref{fig:forward-reverse-process}, we illustrate this observation by showing both the forward and reverse process for one of our SDEs \eqref{eq:ouve-sde}. Without the score term, the state $\x_t$ of the reverse process would diverge away from $\y$. The score model can therefore be interpreted as a learned control strategy adapted to the problem and dataset, which must steer $\x_t$ towards a plausible sample $\x_0$ in the presence of this repulsive action away from $\y$ due to $-\f$. A general connection between diffusion-based generative modeling and stochastic optimal control has recently been discussed in~\cite{berner2022optimal}.

\begin{figure}[ht]
    \centering
    \includegraphics[width=\linewidth]{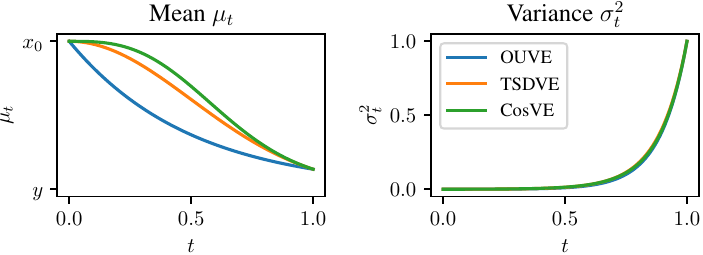}
    \caption{Mean and variance curves of the forward processes described by the \acp{sde} \labelcref{eq:ouve-sde,eq:tsdve-sde,eq:cosve-sde}. The mean curves exhibit different shapes between $\x_0$ and $\y$ but start and end at the same points. The variance curves are all drawn but indistinguishable.}
    \label{fig:means-and-variances}
\end{figure}

\subsection{Three considered SDEs}\label{sec:considered-sdes}
In this work, we consider three \acp{sde} from our family~\eqref{eq:forward-drift-sde-family}:
\begin{align}
    F(t) &= 1 \label{eq:ouve-sde}\,,\\
    F(t) &= t \label{eq:tsdve-sde}\,,\\
    F(t) &= 1 - \cos(\pi t) \label{eq:cosve-sde}\,,
\end{align}
defined on the interval $t \in [0, 1]$. The choice \eqref{eq:ouve-sde} was previously proposed by us in the speech processing literature \cite{welker2022speech}, and we refer to it as the \emph{\aclu{ouve}} (\acs{ouve}) \ac{sde}. We newly propose \eqref{eq:tsdve-sde} and \eqref{eq:cosve-sde} here. We name \eqref{eq:tsdve-sde} the \emph{\aclu{tsdve}} (\acs{tsdve}) SDE due to the decay towards $\y$ depending on $t^2$, see \eqref{eq:mu-tsdve}, and name \eqref{eq:cosve-sde} the CosVE \ac{sde} due to the cosine-based drift expression which we use since a cosine schedule performed best in related work \cite{luo2023image}. To allow for efficient forward sampling in order to perform denoising score matching \cite{song2021sde}, we determine closed-form expressions of the mean $\bmu_t$ and variance $\sigma_t$ of the Gaussian processes described by each SDE. Such closed forms are often not available for general SDEs but can be determined in our case since the drift terms we choose are affine functions of the state $\x_t$, which we do via \cite[eq. (5.50) and (5.53)]{sarkka2019sde}.

\subsubsection{Mean expressions}

The closed-form mean expressions of our SDEs are:
\begin{align}
    \label{eq:mu-ouve}
    \bmu_t^{\text{OUVE}} &= e^{-\gamma t} \x_0 + (1 - e^{-\gamma t}) \y\,,\\
    \label{eq:mu-tsdve}
    \bmu_t^{\text{TSDVE}} &= e^{-\gamma \frac{t^2}{2}} \x_0 + (1 - e^{-\gamma \frac{t^2}{2}}) \y\,,\\
    \label{eq:mu-cosve}
    \bmu_t^{\text{CosVE}} &= e^{-\gamma \left(t - \frac{\sin(\pi t)}{\pi}\right)} \x_0 + \left(1-e^{-\gamma \left(t - \frac{\sin(\pi t)}{\pi}\right)}\right) \y\,.
\end{align}
All three expressions describe a linear (pixel-wise) interpolation between $\x_0$ and $\y$, with the interpolation parameter controlled by an exponential (\acs{ouve}), half-Gaussian-shaped (\acs{tsdve}), or more complex (CosVE) decay over time $t$. \cref{fig:forward-reverse-process} illustrates the decay of the mean and simultaneous addition of noise for the \ac{ouve} \ac{sde}.

For all \acp{sde}, $\bmu_t \neq \y$ for all finite $t$, which may seem like an issue since the aim was to have the process move towards $\y$. However, letting $\z \sim \mathcal{N}(0, I)$, it is only required that the distributions of $(\bmu_T + \sigma_T\z)$ and $(\y + \sigma_T\z)$ are close, so that the latter can function as a plausible initial value for the reverse sampling process. We can control how well the distributions of these two expressions match, either by increasing the stiffness $\gamma$ at the cost of potentially destabilizing the reverse process, or by increasing $\sigma\submax$ to further smooth the density functions of both distributions at the cost of more reverse iterations.

\subsubsection{Variance expressions}
\label{sec:numerical-forward-sde}
The closed-form expression for the variance can be complex and may not be possible to solve depending on the particular choices of the drift and diffusion coefficient. For the OUVE and TSDVE SDEs, we let the software \emph{Mathematica} determine closed forms of the variance, but we could not retrieve a closed-form solution for the CosVE SDE. However, we know that the variance must be independent of $\x_t$ and $\y$ due to the linearity of our forward SDEs, and only depends on $t$ and $\gamma$. We can thus use numerical integration to approximately solve for the variance at a discrete set of points $P \subset [t_\varepsilon, T]$ and, based on those, construct an interpolator for arbitrary $t \in [t_\varepsilon, T]$. The values $\{\sigma_p \mid p\in P\}$ must only be computed once as they are independent of $\x_t$ and $\y$. This simple observation allows us to more freely choose combinations of $F(t)$ and $g(t)$, thus circumventing the problem of an overly smooth variance noted in~\cite{luo2023image} which occurs for their choice of coupled drift and diffusion coefficients, and enabling future exploration of other \acp{sde}. For CosVE, we use 1000 equidistant points on the interval $[t_\varepsilon, T]$ as $P$, \texttt{scipy.integrate.quad} with default settings for numerical integration, and subsequently use cubic spline interpolation.

\begin{table*}[ht]
    \begin{subtable}[h]{0.35\textwidth}
        \centering
        \begin{tabular}{lcccccc}
            \toprule
            SDE & $\gamma$ & $\sigma\submax$ & $\sigma\submin$ & $t_\varepsilon$ & N
            \\\midrule
            OUVE & 1 & 1.0 & 0.01 & 0.03 & 100\\
            TSDVE & 2 & 1.0 & 0.01 & 0.03 & 100\\
            CosVE & 1 & 1.0 & 0.01 & 0.03 & 100\\
            \bottomrule
        \end{tabular}
        \caption{Process parameterization}
        \label{tab:process-params}
    \end{subtable}
    \hfill
    \begin{subtable}[h]{0.25\textwidth}
        \centering
        \begin{tabular}{lc}
            \toprule
            Learning rate & $10^{-4}$\\
            Batch size per GPU & 8\\
            Number of GPUs & 4\\
            Max. epochs & 300\\
            \bottomrule
        \end{tabular}
        \caption{Training hyperparameters}
        \label{tab:training-params}
    \end{subtable}
    \hfill
    \begin{subtable}[h]{0.3\textwidth}
        \centering
        \begin{tabular}{lr}
            \toprule
            DriftRec & 65.6\,M\\
            Regression Baseline & 65.5\,M\\
            I2SB~\cite{liu2023i2sb} & 65.6\,M\\
            QGAC~\cite{ehrlich2020quantization} & 259.4\,M\\
            FBCNN~\cite{jiang2021towards} & 71.9\,M\\
            DDRM~\cite{kawar2022jpeg} (CelebA-HQ) & 113.7\,M\\
            \bottomrule
        \end{tabular}
        \caption{Number of DNN parameters.}
        \label{tab:model-size}
    \end{subtable}
    \caption{Parameters for the \acp{sde} (a) and network training (b), and model size in number of parameters (c).}
    \label{tab:params}
\end{table*}

\subsection{Architecture and training procedure}\label{sec:architecture-and-training}

We utilize the NCSN++ architecture \cite{song2021sde} both for the regression baseline and as a score network and use the same-sized network for all training runs and datasets. We train the regression baseline $R_\theta$ to recover $\x_0$ given $\y$ via an $L_2$ loss:
\begin{equation}
    \theta^* = \argmin_\theta \mathbb{E}_{t,(\x_0,\y)} \left[ \norm{
        R_\theta(\y) - \x_0
    }_2^2 \right]\,,
\end{equation}
where we pass a constant dummy value of $t=1$ to the time embedding layers~\cite{song2021sde} of NCSN++ to avoid making any changes to the DNN that may affect the qualitative behavior of each layer.
To train the score models $S_\theta$, we use the idea that diffusion models based on our linear \acp{sde} can be trained by the same Denoising Score Matching~\cite{vincent2011connection} target as in \cite{song2021sde}, but conditioned on $\y$:
\begin{equation}\label{eq:dsm-main}
\begin{split}
    \theta^* = &\argmin_\theta \mathbb{E}_{t,(\x_0,\y),\z,\x_t} \left[\phantom{\norm{\grad{\x_t}}_2^2} \right.
    \\
    &\left. \hspace{1em} \norm{
        S_\theta(\x_t, \y, t) - \grad_{\x_t} \log p_{0t}(\x_t | \x_0, \y)
    }_2^2 \right]\,,
\end{split}
\end{equation}
where $p_{0t}(\x_t | \x_0, \y) = \mathcal{N}(\bmu_t, \sigma_t^2 \mathbf{I})$ is the so-called \emph{perturbation kernel}, i.e., the distribution of the forward process at process time $t$. Here, $\mathcal{N}(\cdot, \cdot)$ denotes a Gaussian distribution, and $\mathbf I$ is the identity matrix matching the total dimensionality of $\x_0$. The only changes in the objective~\eqref{eq:dsm-main} over the usual Denoising Score Matching objective are that the expectation is also calculated over $\y$ and that $\y$ is provided to the score network. We provide $\y$ as an input to $S_\theta$ by concatenating $\x_t$ and $\y$ along the channel dimension, for which we change the number of input channels of NCSN++ from 3 to 6, which increases the number of DNN parameters by only about $0.1\,\%$ (see \cref{tab:model-size}).

Due to $p_{0t}$ being a Gaussian distribution, a closed form of the score term can be derived as:
\begin{equation}
    \grad_{\x_t} \log p_{0t}(\x_t|\x_0, \y) = \frac{-(\x_t - \bmu_t)}{\sigma_t^2}
\end{equation}
which allows to write the training objective \eqref{eq:dsm-main} as:
\begin{equation}\label{eq:dsm-almost-easy}
    \theta^* = \argmin_\theta \mathbb{E}_{t,(\x_0,\y),\z,\x_t} \left[ \norm{
        S_\theta(\x_t, \y, t) + \left(\sfrac{\z}{\sigma_t}\right)
    }_2^2 \right]\,,
\end{equation}
\begin{equation}
    \text{where}\hspace{1.5em} \x_t := \bmu_t + \sigma_t \z\,, \hspace{1.5em}\z \sim \mathcal{N}(0, \mathbf I)\,,
\end{equation}
with $(\x_0, \y)$ being clean/corrupted image pairs sampled from the dataset. We found that training with this objective can be unstable due to very small values of the denominator $\sigma_t$ when $t$ is close to zero. Since we have $\sigma_t$ available in closed form at the time of sampling and do not need to learn it, similarly to~\cite{song2021sde} we use a slight modification of the objective:
\begin{equation}\label{eq:dsm-easy}
    \theta^* = \argmin_\theta \mathbb{E}_{t,(\x_0,\y),\z,\x_t} \left[ \norm{
        \tilde{S}_\theta(\x_t, \y, t) + \z
    }_2^2 \right]\,,
\end{equation}
with $\tilde{S}_\theta := \frac{S_\theta}{\sigma_t}$, which can also be interpreted as weighting \eqref{eq:dsm-almost-easy} by $\frac{1}{\sigma_t^2}$. This objective has the advantageous effect that the target output of the DNN, $\z$, has zero mean and unit variance for all process times $t$.

\section{Experiments}

\begin{figure*}[ht]
    \centering
    \includegraphics[width=\textwidth]{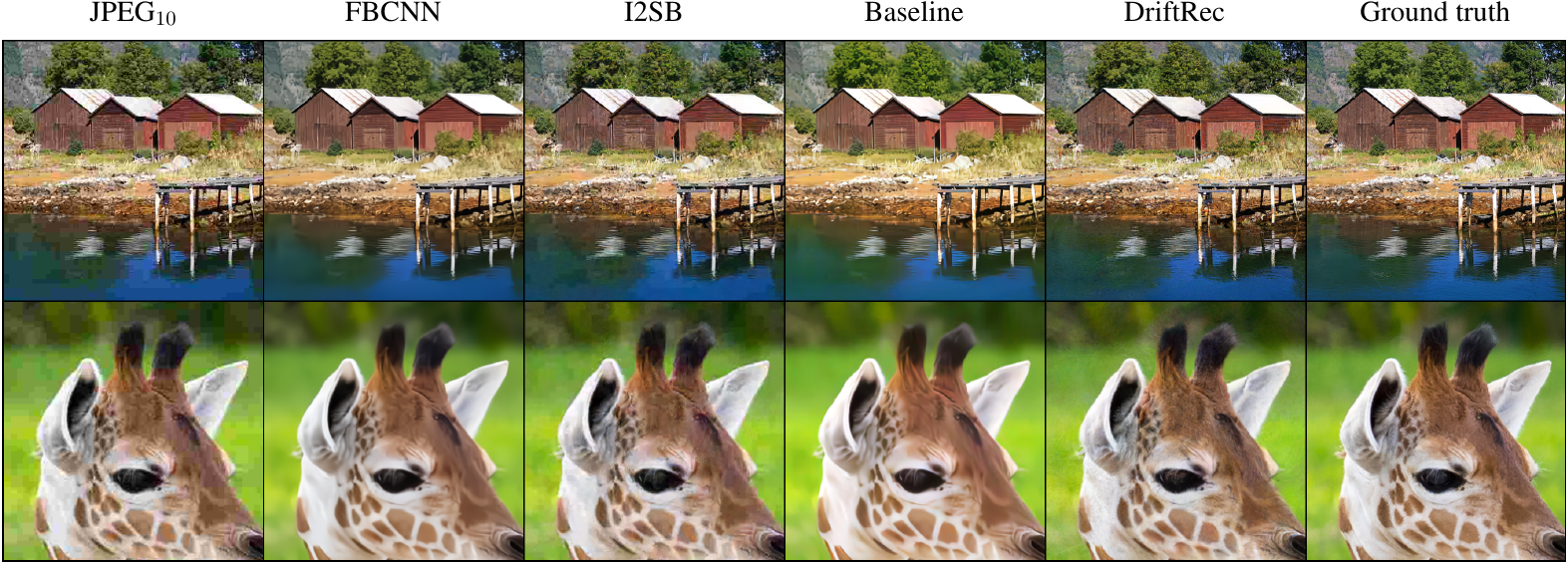}
    \caption{Example reconstructions on the D2KF2K test set for JPEG quality factor 10, comparing FBCNN~\cite{jiang2021towards}, I2SB~\cite{liu2023i2sb}, the regression baseline, and our proposed method DriftRec. Reconstructions by FBCNN and the baseline exhibit a blurry, painting-like look, whereas DriftRec generates high-frequency detail better matching that of the ground-truth image. I2SB leaves significant blocky artifacts in this blind setting. Best viewed zoomed in.}
    \label{fig:images-d2kf2k-jpeg10}
\end{figure*}

\subsection{Datasets}
We use three datasets for our experiments: \B{(1)} \celeba{}, the CelebA-HQ dataset \cite{karras2018progressive} resized to 256x256. \B{(2)} a composite dataset (\enquote{D2KF2K}) consisting of the DIV2K and Flickr2K datasets~\cite{agustsson2017div2k,timofte2017flickr2k} with 3550 total images. We randomly split each dataset into 80\% training, 5\% validation, and 15\% test data. \B{(3)} The LIVE1 dataset~\cite{sheikh2006statistical}, which we use in its entirety as a test set for smaller evaluation experiments. For training, we let $\x_0$ be random $256\times{}256$ patches of the clean images from the training dataset, using uniformly sampled random resized crops as a data augmentation. To generate $\y$, we sample a JPEG \aclu{qf} uniformly at random from 0 to 100 for each image $\x_0$ and training step, and then apply JPEG compression to $\x_0$ using the \texttt{Pillow} Python library~\cite{clark2015pillow}. During evaluation, we set the JPEG \acp{qf} to a constant value across all compared images and models. We do not provide \acp{qf} to our models, so they must perform the reconstruction while being \emph{blind} both to the amount and the internals of the used JPEG compression.

\subsection{Training and process parameterization}
For training, we use the \emph{AdamW} optimizer~\cite{loshchilov2018decoupled}, and set the hyperparameters of the process and training as listed in \Cref{tab:params}. We chose $\gamma$ for the \ac{ouve} and CosVE \acp{sde} via manual grid search in $\{1, 1.5, 2\}$, finding $\gamma=1$ to generally perform best. We set $\gamma_{\text{TSDVE}} = 2\gamma$ for the \ac{tsdve} \ac{sde} to match the means of the terminal distributions $(t=T)$, see \cref{eq:mu-ouve,eq:mu-tsdve}. $\sigma\submax$ was empirically chosen by finding a value that ensures that $(\bmu_T+\sigma_T \z)$ and $(\y + \sigma_T \z)$ are visually indistinguishable. Similarly, we determined $\sigma\submin = 0.01$ so that the closed-form variance $\sigma_t$ is small enough at $t=t_\varepsilon$ so that the noise is not visible. We linearly rescale all clean and corrupted input images from RGB values of $[0, 255]$ to $[0, 1]$, and the process parameterization above should be interpreted relative to this. Similarly to~\cite{song2021sde}, we update an exponential moving average with a decay of 0.999 for all network parameters after each training step and use these parameters for evaluation. We always train separate models for the two datasets D2KF2K and \celeba{}.

\subsection{Compared methods}
\label{sec:compared-methods}
Besides our regression baseline that uses the same network architecture, we also compare against state-of-the-art methods for the JPEG restoration task, namely \emph{\aclu{qgac}} (\acs{qgac})~\cite{ehrlich2020quantization}, which utilizes the quantization matrices stored in JPEG files to adapt to quantization levels and it thus non-blind, and \emph{\aclu{fbcnn}} (\acs{fbcnn})~\cite{jiang2021towards}, which relies on learned quality factor prediction. We also compare against the diffusion-based methods DDRM~\cite{kawar2022denoising}, as adapted for JPEG restoration in~\cite{kawar2022jpeg}, and I2SB~\cite{liu2023i2sb}, see~\cref{sec:related-work}. For the closely related I2SB, in order to perform a fair comparison, here we align the network, training and sampling to our DriftRec models as much as possible: We use the publicly available code\footnote{\url{https://github.com/NVlabs/I2SB}} with recommended parameters and conditioning on $\y$ (option \enquote{\texttt{--cond-x1}}), but change the configuration to use the same score network (NCSN++), training dataset (D2KF2K), blind training scenario (uniformly random \ac{qf} 0--100), total amount of observed training samples ($\sim 8\times10^{6}$), and DNN evaluations during sampling (100) as our DriftRec models.

\section{Results and Discussion}
\label{sec:results-and-discussion}

\begin{figure*}
    \centering
    \includegraphics[width=\textwidth]{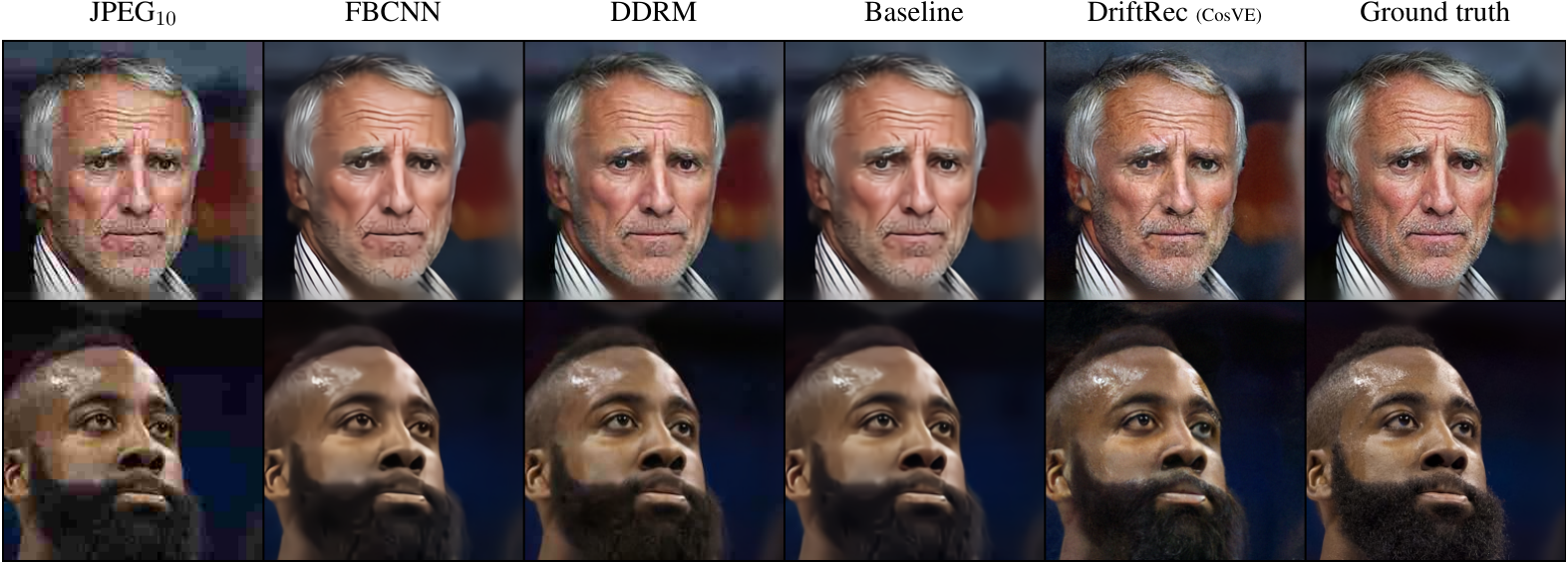}
    \caption{Example reconstructions on \celeba{} for JPEG quality factor 10, comparing FBCNN~\cite{jiang2021towards}, \ac{ddrm}~\cite{kawar2022jpeg} and the regression baseline against our method. Best viewed zoomed in.}
    \label{fig:images-celebahq-jpeg10}
\end{figure*}

For all experiments besides those in \cref{sec:sampler-experiments}, we generate reconstructions from our DriftRec models with the Euler-Maruyama sampler~\cite{kloeden2011numerical} using $N=100$ uniform discretization steps. For fairness, for DDRM~\cite{kawar2022jpeg}, we follow the authors' method but retrieve single samples rather than a multi-sample average, as in our method. Also for a fair comparison, for I2SB~\cite{liu2023i2sb}, we use the authors' default settings but set the number of DNN evaluations to 100, as in our method. For FBCNN~\cite{jiang2021towards}, QGAC~\cite{ehrlich2020quantization}, and the regression baseline, we retrieve reconstructions via a single forward pass. We use the publicly available codes and checkpoints of DDRM, QGAC, and FBCNN. For QGAC on \celeba{}, we also evaluate the GAN weights with an interpolation parameter of $\alpha = 0.7$, as in~\cite{ehrlich2020quantization} (\enquote{QGAC$_\text{GAN}$}).

\subsection{Qualitative evaluation}
In \Cref{fig:images-d2kf2k-jpeg10}, we show example reconstructions on our D2KF2K test set for \acf{qf} 10, comparing \ac{fbcnn} and the regression baseline against DriftRec with the \ac{ouve} \ac{sde}. Qualitatively, both \ac{fbcnn} and the regression baseline reconstruct major features well but fail to produce natural high-frequency detail. This is particularly visible for hair, water, and foliage textures, and results in a painting-like look. In contrast, our approach reconstructs more perceptually natural images and shows a remarkable ability to reconstruct natural textures that are plausible given the corrupted image. In \Cref{fig:images-celebahq-jpeg10}, we show reconstructions for \celeba{} and additionally compare against \ac{ddrm}. Interestingly, even though \ac{ddrm} is also diffusion-based and trained on the same dataset, it produces noticeably less high-frequency detail than our method and leaves some blocky artifacts in the reconstruction.

\subsection{Quantitative evaluation}
\begin{table*}
    \begin{subtable}[h]{0.47\textwidth}
        \centering
        \begin{tabular}{lrrrrrr}
        \toprule
        {} &  SSIM$^\uparrow$ & PSNR$^\uparrow$ & LPIPS$^\downarrow$ & KID$^\downarrow$ & FID$^\downarrow$ \\
        \midrule
        JPEG$_{30}$ & 0.90 & 31.85 & 7.5 & 2.42 & 34.1 \\
        $^\ddag$QGAC & 0.94 & 33.87 & 8.6 & 2.45 & 34.0 \\
        $^\ddag$QGAC$_{\text{GAN}}$ & 0.93 & 33.49 & 5.3 & 0.55 & 17.5 \\
        FBCNN & \B{0.94} & \B{34.18} & 7.9 & 2.37 & 32.8 \\
        $^\ddag$DDRM & 0.92 & 32.57 & 6.2 & 1.85 & 28.5 \\
        Baseline & \B{0.94} & 33.95 & 7.7 & 2.64 & 34.2 \\
        DriftRec / OUVE & 0.89 & 31.57 & 4.1 & 0.54 & 17.3 \\
        DriftRec / TSDVE & 0.88 & 31.46 & 4.5 & 0.80 & 20.0 \\
        DriftRec / CosVE & 0.88 & 31.44 & \B{3.9} & \B{0.44} & \B{16.5} \\
        \bottomrule
        \end{tabular}
        \caption{JPEG Quality Factor \textbf{QF = 30}}
        \label{tab:metrics:celeba-30}
    \end{subtable}
    \hfill
    \begin{subtable}[h]{0.47\textwidth}
        \centering
        \begin{tabular}{lrrrrrr}
        \toprule
        {} &  SSIM$^\uparrow$ & PSNR$^\uparrow$ & LPIPS$^\downarrow$ & KID$^\downarrow$ & FID$^\downarrow$ \\
        \midrule
        JPEG$_{20}$ & 0.88 & 30.49 & 11.2 & 2.92 & 39.6 \\
        $^\ddag$QGAC & 0.92 & 32.64 & 11.3 & 3.15 & 41.0 \\
        $^\ddag$QGAC$_{\text{GAN}}$ & 0.91 & 32.28 & 7.5 & 0.75 & 20.0 \\
        FBCNN & \B{0.92} & \B{32.92} & 10.7 & 3.09 & 39.9 \\
        $^\ddag$DDRM & 0.90 & 31.66 & 8.0 & 2.03 & 30.5 \\
        Baseline & \B{0.92} & 32.84 & 10.4 & 3.32 & 40.7 \\
        DriftRec / OUVE & 0.86 & 30.45 & 5.7 & 0.67 & 19.3 \\
        DriftRec / TSDVE & 0.86 & 30.40 & 6.0 & 0.88 & 21.5 \\
        DriftRec / CosVE & 0.86 & 30.37 & \B{5.2} & \B{0.50} & \B{17.8} \\
        \bottomrule
        \end{tabular}
        \caption{JPEG Quality Factor \textbf{QF = 20}}
        \label{tab:metrics:celeba-20}
    \end{subtable}

    \vspace{1em}
    \begin{subtable}[h]{0.47\textwidth}
        \centering
        \begin{tabular}{lrrrrrr}
        \toprule
        {} &  SSIM$^\uparrow$ & PSNR$^\uparrow$ & LPIPS$^\downarrow$ & KID$^\downarrow$ & FID$^\downarrow$ \\
        \midrule
        JPEG$_{10}$ & 0.81 & 27.94 & 21.8 & 4.26 & 53.7 \\
        $^\ddag$QGAC & 0.88 & 30.43 & 17.0 & 4.61 & 55.4 \\
        $^\ddag$QGAC$_{\text{GAN}}$ & 0.87 & 30.10 & 12.2 & 1.49 & 27.8 \\
        FBCNN & 0.88 & 30.67 & 16.5 & 4.41 & 53.0 \\
        $^\ddag$DDRM & 0.86 & 29.98 & 11.4 & 2.24 & 33.4 \\
        Baseline & \B{0.89} & \B{30.80} & 15.7 & 4.38 & 51.2 \\
        DriftRec / OUVE & 0.81 & 28.34 & 9.7 & 0.97 & 23.4 \\
        DriftRec / TSDVE & 0.81 & 28.37 & 9.7 & 1.11 & 25.0 \\
        DriftRec / CosVE & 0.81 & 28.36 & \B{8.7} & \B{0.68} & \B{20.8} \\
        \bottomrule
        \end{tabular}
        \caption{JPEG Quality Factor \textbf{QF = 10}}
        \label{tab:metrics:celeba-10}
    \end{subtable}
    \hfill
    \begin{subtable}[h]{0.47\textwidth}
        \centering
        \begin{tabular}{lrrrrrr}
        \toprule
        {} &  SSIM$^\uparrow$ & PSNR$^\uparrow$ & LPIPS$^\downarrow$ & KID$^\downarrow$ & FID$^\downarrow$ \\
        \midrule
        JPEG$_{5}$ & 0.70 & 24.87 & 38.6 & 7.66 & 88.8 \\
        $^\ddag$QGAC & 0.73 & 25.55 & 29.1 & 10.22 & 104.2 \\
        $^\ddag$QGAC$_{\text{GAN}}$ & 0.67 & 24.16 & 37.0 & 12.07 & 118.5 \\
        FBCNN & 0.81 & 27.41 & 23.9 & 7.90 & 85.6 \\
        $^\ddag$DDRM & 0.81 & 27.89 & 15.8 & 2.40 & 36.5 \\        
        Baseline & \B{0.83} & \B{28.34} & 22.3 & 5.43 & 62.0 \\
        DriftRec / OUVE & 0.73 & 25.78 & 15.6 & 1.44 & 29.7 \\
        DriftRec / TSDVE & 0.73 & 25.79 & 16.0 & 1.58 & 31.6 \\
        DriftRec / CosVE & 0.73 & 25.82 & \B{14.5} & \B{1.09} & \B{26.4} \\
        \bottomrule
        \end{tabular}
        \caption{JPEG Quality Factor \textbf{QF = 5}}
        \label{tab:metrics:celeba-5}
    \end{subtable}
    \caption{Distribution-based metrics (KID~\cite{binkowski2018demystifying}, FID~\cite{FIDref}), SSIM~\cite{wang2004image}, PSNR, and the learned perceptual metric LPIPS~\cite{zhang2018perceptual} (\enquote{LP}) for 4500 \celeba{} test images, comparing reconstructed images of all methods to the compressed images and the ground truth for JPEG \acp{qf} (a) 30, (b) 20, (c) 10 and (d) 5. KID and LPIPS scores are multiplied by 100 for readability. Arrows indicate whether low or high values are better. Best values in bold. All DriftRec variants achieve significantly better perceptual (LPIPS) and distributional (KID, FID) scores than all other methods. $^\ddag$: non-blind method.}
    \label{tab:metrics}
\end{table*}

In \cref{tab:metrics}, we quantitatively compare all methods on \celeba{} for the quality factors \ac{qf} $\in \{30, 20, 10, 5\}$, using the reference-based metrics \ac{ssim} \cite{wang2004image} and \ac{psnr}, the perceptual metric \acf{lpips} \cite{zhang2018perceptual}, and the distribution-based metrics \ac{fid} \cite{FIDref} and \ac{kid} \cite{binkowski2018demystifying}. We use the official code by Zhang et al.~\cite{zhang2018perceptual} to calculate \ac{lpips}, the \texttt{scikit-image} Python library~\cite{van2014scikit} to calculate \ac{ssim} and \ac{psnr}, and the \texttt{piq} Python library~\cite{kastryulin2022piq} to calculate \ac{kid} and \ac{fid}, using default settings throughout. We evaluate all metrics on our test set of 4500 \celeba{} images.

Judging from \ac{kid} and \ac{fid}, all variants of DriftRec consistently model the clean image distribution of \celeba{} significantly more faithfully than the compared methods. All other methods, except the also diffusion-based DDRM~\cite{kawar2022jpeg}, either worsen \ac{kid} and \ac{fid} scores or only marginally improve them compared to the compressed images. The GAN-based variant of QGAC is still outperformed by DriftRec in this regard, despite being generative, non-blind, and using a larger network (see \cref{tab:params}). It is important to point out that \ac{kid} and \ac{fid} only compare the \emph{distributions} of extracted features from all test images against the features from the ground-truth distribution, and thus only judge how plausible the reconstructions are overall. They cannot be used directly to make a statement about the faithfulness of each single reconstruction to its ground-truth reference image. However, the reference-based perceptual metric \ac{lpips} also shows a consistent advantage of all DriftRec variants methods in terms of perceptual quality for all \ac{qf} values, which matches our subjective visual inspection.

For \ac{ssim} and \ac{psnr}, on the other hand, the compared methods generally achieve improvements throughout while DriftRec does not. Our regression baseline achieves strong performance in this regard and overall performs rather similarly to, and overall slightly better than, \ac{qgac} and \ac{fbcnn}. The low \ac{psnr} and \ac{ssim} improvements of DriftRec are to be expected due to the generative and probabilistic nature of the method. DriftRec always attempts to generate high-frequency detail, even for low \ac{qf} where such details must be generated from very little information. The compared approaches resort to reconstructing blurry images, which however achieve decent \ac{psnr} and \ac{ssim} scores. As noted in~\cite{dahl2017pixel}, generated high-frequency content is generally disincentivized by classic reference-based metrics such as \ac{psnr} and \ac{ssim}. \ac{ssim} has also been shown to not correlate well with human perceptual preference~\cite{dahl2017pixel,ledig2017photo}, with \ac{lpips} scores being aligned better~\cite{mier2021deep}. The general phenomenon of this \emph{perception-distortion tradeoff} has been discussed at length in~\cite{blau2018perception}. With our method DriftRec, we have thus emphasized significantly better \emph{perception} at the cost of more \emph{distortion}. We argue that this is a useful direction for JPEG restoration as the application of key interest here is the enhancement of compressed natural images, for which it is usually of higher importance to be visually pleasing and natural-looking rather than to be faithful in a per-pixel sense.

In \cref{tab:metrics:d2kf2k}, we show metrics for our D2KF2K test set on \acp{qf} 10, 20, and 30. Here we omit the distribution-based metrics KID and FID due to the test set being comparatively small with only 498 test images but additionally show \psnrb{}~\cite{tadala2012psnrb}, which combines PSNR with the \ac{bef}~\cite{tadala2012psnrb}, a reference-free measure of blockiness whose raw values we also list. Overall, we draw the same conclusions as for our evaluation on the \celeba{} test set: DriftRec achieves the best perceptual quality with a margin but does not improve PSNR over the compressed images on average. We note however that DriftRec does improve \psnrb{} over the compressed images. When considering only the blockiness measure \ac{bef} that is part of \psnrb{}, we find that all DriftRec models exhibit a better reduction of blockiness than all other methods, which can also be corroborated by visual inspection of the example images in \cref{fig:images-d2kf2k-jpeg10,fig:images-celebahq-jpeg10}. Comparing against the closely related method I2SB \cite{liu2023i2sb}, using the same score network architecture and training dataset -- see \cref{sec:compared-methods} for details -- we find that DriftRec also achieves significantly better LPIPS and BEF, with I2SB performing worse than all other methods for deblocking (BEF). Referring back to \cref{fig:images-d2kf2k-jpeg10}, we argue that DriftRec clearly reconstructs more perceptually natural and deblocked images than I2SB. Since the I2SB paper \cite{liu2023i2sb} shows highly competitive results when training on a single QF (non-blind) with a larger network, this may indicate that our method is more training-efficient or more well-suited as a generalist blind JPEG restoration model than I2SB.

Regarding the choice of \ac{sde} within our framework, we find only minor differences in performance. For the \celeba{} dataset, the CosVE SDE \eqref{eq:cosve-sde} seems to perform best, whereas for D2KF2K the OUVE SDE \eqref{eq:ouve-sde} achieves the best results. The TSDVE SDE \eqref{eq:tsdve-sde} in both cases performs somewhere in between the other two. This is, at first glance, at odds with the observations made by Luo et al.~\cite{luo2023image} where the authors show a larger quality difference between the three SDEs they compare, which have similar characteristics to our SDEs (constant, linear and cosine schedules). We note, however, that the SDEs compared in~\cite{luo2023image} differ from ours in the diffusion coefficient $g(t)$, and describe three forward processes whose variance evolves differently to each other. In contrast, our \acp{sde} in \cref{eq:ouve-sde,eq:tsdve-sde,eq:cosve-sde} all share the same diffusion coefficient $g(t)$ -- resulting in highly similar variance evolutions to each other -- and effectively differ only in the evolution of the process mean, see \cref{fig:means-and-variances}. This suggests that the variance schedule is of much higher importance than the mean schedule for reconstruction quality. The optimal choice of the SDE for any given task is currently unknown, which warrants further research.

\begin{table*}
    \centering
    \begin{tabular}{lrrrrr|rrrrr|rrrrr}
        \toprule
        {} & \multicolumn{5}{c|}{QF = {30}}
            & \multicolumn{5}{c|}{QF = {20}}
            & \multicolumn{5}{c}{QF = 10} \\
        \cmidrule{2-16}
        {} &  SSIM & PSNR & {\footnotesize PSNR}B & LP & BEF
            &  SSIM & PSNR & {\footnotesize PSNR}B & LP & BEF
            &  SSIM & PSNR & {\footnotesize PSNR}B & LP & BEF\\
        \midrule
        JPEG & 0.90 & 30.82 & 28.99 & 14.8 & 17.4
             & 0.87 & 29.46 & 27.59 & 20.4 & 24.2
             & 0.80 & 26.91 & 25.14 & 32.5 & 39.5\\
        $^\ddag$QGAC & \B{0.93} & 32.71 & 32.36 & 14.5 & 3.3
             & \B{0.91} & 31.44 & 31.17 & 18.0 & 3.6
             & \B{0.86} & 29.11 & 28.96 & 25.2 & 3.7\\
        FBCNN & \B{0.93} & \B{32.96} & \B{32.53} & 13.6 & 3.5
             & \B{0.91} & \B{31.65} & \B{31.31} & 17.1 & 3.9
             & \B{0.86} & \B{29.25} & \B{29.28} & 24.6 & 4.0 \\
        Baseline & 0.89 & 26.99 & 27.25 & 15.4 & 3.4
            & 0.87 & 26.42 & 26.62 & 18.8 & 3.4
            & 0.82 & 25.10 & 25.26 & 26.1 & 2.8 \\
        I2SB & 0.89 & 30.55 & 30.29 & 11.5 & 3.4
            & 0.87 & 29.34 & 29.03 & 14.7 & 4.6
            & 0.80 & 26.96 & 26.68 & 23.3 & 6.6 \\
        DriftRec & & & & & & & & & & & & & & &\\
        \quad{}OUVE & 0.86 & 30.08 & 30.07 & \B{8.6} & 0.9
            & 0.84 & 28.85 & 28.83 & \B{11.3} & 1.2
            & 0.74 & 26.20 & 26.22 & \B{19.8} & 1.3\\
        \quad{}TSDVE & 0.85 & 29.59 & 29.63 & 9.7 & \B{0.5}
            & 0.82 & 28.37 & 28.41 & 12.7 & \B{0.6}
            & 0.74 & 25.97 & 26.01 & 20.7 & 0.8 \\
        \quad{}CosVE & 0.85 & 29.56 & 29.60 & 10.3 & \B{0.5}
            & 0.82 & 28.29 & 28.33 & 13.5 & \B{0.6}
            & 0.74 & 25.82 & 25.87 & 21.4 & \B{0.7}\\
        \bottomrule
    \end{tabular}
    \caption{Metrics for the D2KF2K test set (\acp{qf} 30, 20 and 10). Best values in bold. DriftRec consistently achieves the best perceptual values (LPIPS, \enquote{LP} here) values by a large margin. For readability, LPIPS is multiplied by 100, and the blockiness measure BEF~\cite{tadala2012psnrb} by 10,000. Higher is better for SSIM and PSNR(B); lower is better for LPIPS and BEF. $^\ddag$: non-blind method.}
    \label{tab:metrics:d2kf2k}
\end{table*}

\subsection{Unaligned double JPEG compression}
\begin{table*}
    \centering
    \begin{tabular}{lccccc}
    \toprule
        QF$_{1/2}$ & JPEG & $^\ddag$QGAC & FBCNN & Baseline & DriftRec (OUVE)\\
    \midrule
    30/10 &
    0.76\textbar25.4\textbar23.6\textbar31.4 & 0.82\textbar27.4\textbar27.2\textbar26.9 & 0.82\textbar27.5\textbar27.2\textbar26.0 & 0.81\textbar26.8\textbar26.6\textbar25.9 & 0.71\textbar24.9\textbar24.8\textbar17.3 \\
    10/30 &
    0.76\textbar25.6\textbar25.0\textbar31.9 & 0.79\textbar26.4\textbar26.4\textbar32.5 & 0.79\textbar26.4\textbar26.4\textbar31.3 & 0.78\textbar26.0\textbar25.9\textbar29.8 & 0.73\textbar25.3\textbar25.3\textbar19.2 \\
    \midrule
    30 & 0.88\textbar29.4\textbar27.4\textbar12.9 & 0.91\textbar31.2\textbar30.7\textbar13.0 & 0.91\textbar31.4\textbar30.8\textbar11.9 & 0.91\textbar29.7\textbar29.3\textbar11.8 & 0.84\textbar28.4\textbar28.3\textbar\phantom{0}7.8 \\
    30/30 &
    0.85\textbar28.0\textbar26.6\textbar17.7 & 0.88\textbar29.7\textbar29.4\textbar17.6 & 0.88\textbar29.9\textbar29.5\textbar16.3 & 0.88\textbar28.8\textbar28.6\textbar15.6 & 0.81\textbar27.5\textbar27.5\textbar\phantom{0}9.5 \\
    \midrule
    50/30 &
    0.86\textbar28.6\textbar26.9\textbar15.4 & 0.90\textbar30.4\textbar30.0\textbar15.2 & 0.90\textbar30.7\textbar30.1\textbar13.8 & 0.89\textbar29.4\textbar29.0\textbar13.5 & 0.83\textbar27.9\textbar27.9\textbar\phantom{0}8.4 \\
    30/50 &
    0.86\textbar28.6\textbar27.7\textbar15.6 & 0.89\textbar30.0\textbar29.7\textbar16.2 & 0.89\textbar30.1\textbar29.9\textbar15.0 & 0.89\textbar29.1\textbar28.9\textbar14.0 & 0.83\textbar28.1\textbar28.1\textbar\phantom{0}8.5 \\
    \bottomrule
    \end{tabular}
    \caption{Results for unaligned double JPEG compression on the LIVE1 dataset. Each entry lists (SSIM\textbar{}PSNR\textbar{}\psnrb{}\textbar{}LPIPS*100). DriftRec exhibits less deterioration in LPIPS than other methods when swapping QF$_1$ and QF$_2$ and achieves the best LPIPS by a large margin, without having been trained for this task. $^\ddag$: non-blind method.}
    \label{tab:metrics:dbljpeg-live1}
\end{table*}

\begin{figure*}
    \centering
    \includegraphics[width=\linewidth]{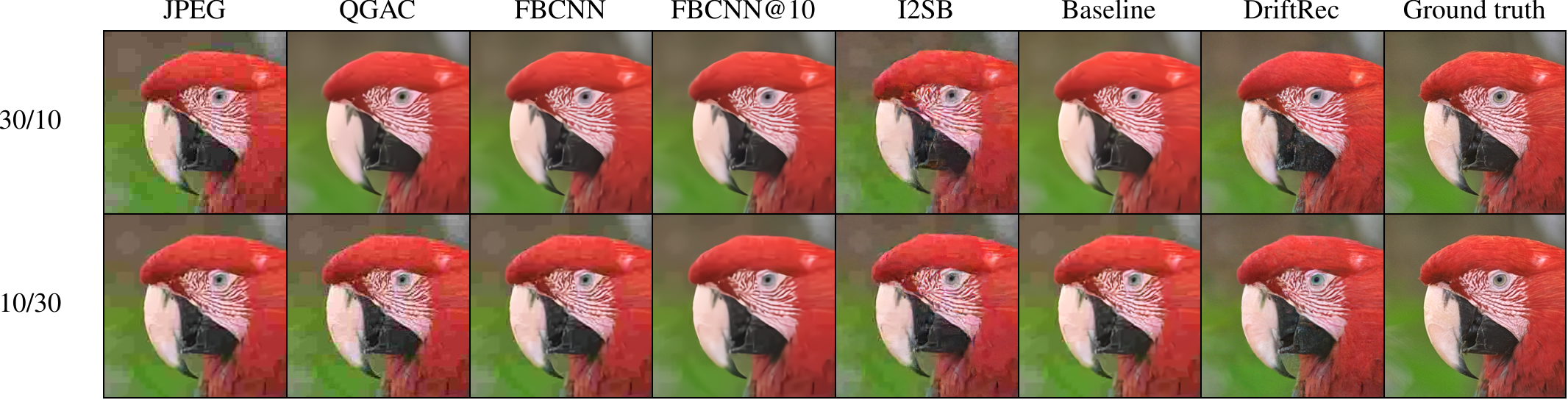}
    \caption{Example images for unaligned double JPEG compression with quality factors \B{(top)} QF$_1$=10, QF$_2$=30 and \B{(bottom)} QF$_1$=30, QF$_2$=10. Only DriftRec achieves reliable blind removal of artifacts and exhibits best perceptual quality. \enquote{FBCNN@10} indicates FBCNN~\cite{jiang2021towards} with a forced (and thus non-blind) quality factor of 10.}
    \label{fig:images-dbljpeg-parrot}
\end{figure*}

Unaligned double JPEG compression is the problem of restoring an image that has been JPEG-compressed twice, with some shifting and/or cropping applied between the first and second compression that makes the second compression unaligned with the $8\times{}8$ blocks of the first JPEG compression. This task is a useful basic model of corruptions encountered in a realistic setting, in particular for images on the internet that may have been edited and compressed multiple times. We follow the scheme used in the FBCNN paper~\cite{jiang2021towards}, where each clean image is compressed with a quality factor QF$_1$, then shifted and cropped by some random amount between 0 and 4 pixels in both directions and compressed again with a quality factor QF$_2$. We report metrics in \cref{tab:metrics:dbljpeg-live1}, and compare example reconstructions in \cref{fig:images-dbljpeg-parrot}. We find that our method achieves more consistent results than other methods when QF$_1$ and QF$_2$ are swapped, in particular in the challenging case of $\text{QF}_2 > \text{QF}_1$, where all other methods fail to effectively remove blocky artifacts, see \cref{fig:images-dbljpeg-parrot}. The perceptual quality when swapping QF$_1$ and QF$_2$, as measured by LPIPS, is especially consistent for our method. Interestingly, DriftRec can achieve this consistent performance without having encountered any doubly compressed image during training and thus shows a remarkable ability to generalize to this task.

For FBCNN~\cite{jiang2021towards}, the authors note that results for double compression can be improved by either (1) manually providing the lowest (dominant) \acl{qf}, (2) automatically finding the dominant \ac{qf} by a complete grid search along all \acp{qf} and possible shifts between (0,0) and (7,7), or (3) adding doubly compressed JPEG images to the training dataset. Since (1) is nonblind and (2) attacks the problem by brute force, for fairness, we do not follow these approaches here. In \cref{fig:images-dbljpeg-parrot}, we show that even the most informed variant (1), \enquote{FBCNN@10}, exhibits worse perceptual quality and higher blockiness than our blind approach. The FBCNN authors found option (3) to perform best, and while we do not employ it here, this simple adjustment to the training process can also be straightforwardly applied for DriftRec. We stress, however, that DriftRec already performs favorably and reliably in this task without employing any of these task-oriented adjustments.

\subsection{Real-world blind restoration}
\begin{figure*}
    \centering
    \includegraphics[width=0.9\textwidth]{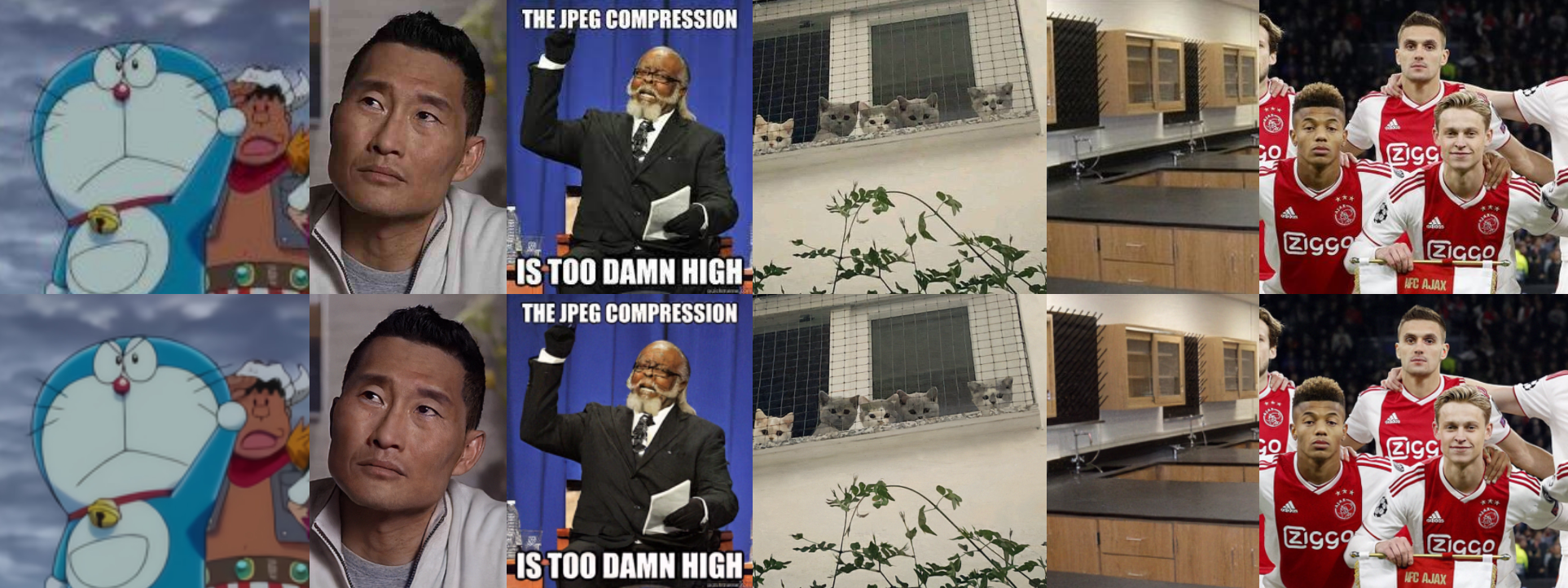}
    \caption{\B{Top:} Real-world JPEG images found on the internet, with quality factors and any other processing being fully unknown. \B{Bottom:} Each image as enhanced by DriftRec (OUVE SDE, trained on D2KF2K). JPEG compression artifacts are removed reliably, and fine textures are preserved or plausibly generated without blurring. Best viewed zoomed in.}
    \label{fig:real-world-jpegs}
\end{figure*}

For practical applicability in a realistic setting, a JPEG restoration model should be able to reconstruct pleasing images in a fully blind setting, where the exact number and amounts of compression, as well as any further processing, is unknown to both the user and the model. In \cref{fig:real-world-jpegs}, we show results from DriftRec for several compressed images found online whose compression settings are unknown and which we provided to DriftRec without any additional information. DriftRec reliably removes block and color artifacts in all images and outputs plausible reconstructions of high perceptual quality. Note that no quantitative statements can be made here due to the uncompressed images being unavailable.

\subsection{Image classification}
To show that our method can also lead to improvements for downstream tasks, in \cref{tab:clf-accuracy}, we report the classifier accuracy (CA) on 5000 images from the ImageNet validation test set, evaluated on the compressed images and their enhanced versions by each method. DriftRec (OUVE, trained on D2KF2K) achieves the best, or close to best, CA improvements. QGAC is not applicable for the 4:4:4 chroma subsampling of ImageNet as it is designed and trained specifically for 4:2:0 subsampling. On re-encoded 4:2:0 images, QGAC with GAN weights mostly achieves the best CA. DriftRec is very similar in performance for \acp{qf} 20 and 30, but in contrast, is not limited to such specific encoding settings and performs fully blind restoration.

\begin{table}
    \centering
    \begin{tabular}{lccc|ccc}
        \toprule
        Subsampling & \multicolumn{3}{c|}{4:4:4} & \multicolumn{3}{c}{4:2:0} \\
        \midrule
        QF & 10 & 20 & 30 & 10 & 20 & 30\\
        \midrule
        JPEG & 65.28 & 73.34 & 76.32 & 63.18 & 70.94 & 74.34 \\
        FBCNN & \underline{66.74} & \underline{74.54} & \underline{76.72} & 65.96 & 73.96 & 76.76 \\
        Baseline & 66.52 & 74.16 & 77.14 & 66.68 & 74.14 & 77.12 \\
        $^\ddag$QGAC & \multicolumn{3}{c|}{--- N/A ---} & 65.26 & 73.16 & 76.18 \\
        $^\ddag$QGAC$_\text{GAN}$ & \multicolumn{3}{c|}{--- N/A ---} & \B{71.50} & \B{75.98} & \underline{77.68} \\
        DriftRec & \B{69.20} & \B{75.88} & \B{77.92} & \underline{68.38} & \underline{75.38} & \B{77.80} \\
        \midrule
        Original & \multicolumn{3}{c|}{--- 80.16 ---} & \multicolumn{3}{c}{--- 79.84 ---}\\
        \bottomrule
    \end{tabular}
    \caption{Classifier accuracy (\%) of ResNet-50 on 5000 ImageNet 256x256 validation images. QGAC does not handle 4:4:4 chroma subsampling and is non-blind. (\B{Best}, \underline{Second}.)}
    \label{tab:clf-accuracy}
\end{table}

\subsection{Sample averaging}
\begin{figure}
    \centering
    \includegraphics[width=\linewidth]{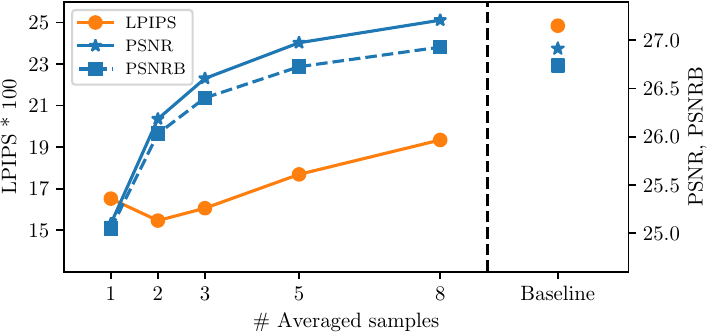}
    \caption{PSNR(B) and LPIPS values when averaging multiple independent samples from a DriftRec model (OUVE, LIVE1, \ac{qf} 10). While averaging more samples consistently improves PSNR(B) and eventually surpasses the regression baseline, doing so also worsens LPIPS for more than 2 samples.}
    \label{fig:sample-averaging-metrics}
\end{figure}

Our method is, like the DDRM-based JPEG restoration method~\cite{kawar2022jpeg}, probabilistic and provides \emph{samples} from an approximated posterior distribution $p(\x_0 | \y)$ rather than single point estimates, and therefore sometimes achieves better and sometimes worse reconstructions (w.r.t. some chosen metric). Kawar et al. propose to stabilize reconstruction performance by averaging 8 samples~\cite{kawar2022jpeg}. Here, we run an experiment in which we average a variable number of samples when reconstructing the LIVE1 dataset~\cite{sheikh2006statistical} from JPEG \ac{qf} 10. The results are shown in \cref{fig:sample-averaging-metrics}. We find that averaging more samples achieves consistently better PSNR(B) scores, but generally worse perceptual \ac{lpips} scores. Interestingly, with 8 averaged samples, our method exhibits a slightly better PSNR(B) score than the regression baseline, while still having a substantially better LPIPS. Varying the number of averaged samples thus allows a flexible choice along the perception-distortion tradeoff~\cite{blau2018perception}. Since drawing samples with DriftRec is trivially parallel, retrieving an optimally averaged reconstruction is also in principle possible on a GPU with no increase in reconstruction time (depending on the available hardware and memory).

\subsection{Sampler settings}
\label{sec:sampler-experiments}
\begin{table}
    \centering
    \begin{tabular}{lrrrrrr}
    \toprule
        Method & SSIM & PSNR & LPIPS & KID & FID & NFE\\
    \midrule
        JPEG$_{10}$ & 0.81 & 27.94 & 21.8 & 4.26 & 53.7 \\
    \midrule
        EuM$\times 100$ & 0.81 & \B{28.36} & \B{8.7} & 0.68 & 20.8 & 100\\
        EuH$\times 100$ & 0.81 & 28.18 & 8.9 & \B{0.39} & \B{18.2} & 200\\
        ODE @ $10^{-1}$ & 0.68 & 26.50 & 23.7 & 3.84 & 47.4 & \B{20}\\
        ODE @ $10^{-2}$ & 0.81 & 28.19 & 10.4 & 1.07 & 24.5 & \B{20}\\
        ODE @ $10^{-3}$ & \B{0.83} & \B{28.37} & 9.2 & 0.78 & 21.0 & 26\\
        ODE @ $10^{-4}$ & 0.82 & 28.20 & 9.1 & 0.57 & 19.5 & 50\\
    \bottomrule
    \end{tabular}
    \caption{Comparison of samplers using the same model, trained with the CosVE SDE (\celeba{}, \ac{qf} 10). \enquote{EuM} is Euler-Maruyama, \enquote{EuH} is Euler-Heun~\cite{roberts2012modify}, %
    and \enquote{ODE} is the RK45 Probability Flow ODE sampler~\cite{song2021sde} with the listed value as both absolute and relative tolerance. \enquote{$\times N$} indicates $N$ discretization steps. NFE is the total number of DNN calls.}
    \label{tab:sampler-settings-celeba-jpeg10-cosve}
\end{table}

Since we derived DriftRec within the continuous \ac{sde}-based formalism for diffusion models, the choice of sampler (numerical solver) for the reverse SDE \eqref{eq:reverse-sde} is flexible after training. Here we evaluate to what extent different samplers result in different metrics, and how many \acf{nfe}, i.e., expensive calls to the score DNN they each require. In \cref{tab:sampler-settings-celeba-jpeg10-cosve}, we compare three samplers on our \celeba{} test set with \ac{qf} 10, using the DriftRec model trained on \celeba{} with the CosVE \ac{sde}: \B{(1)} Euler-Maruyama, see~\cite{song2021sde}, \B{(2)} Euler-Heun adapted for \acp{sde}~\cite{roberts2012modify}, and \B{(3)} four samplers for the Probability Flow ODE~\cite{song2021sde} based on the Dormand-Prince Runge-Kutta scheme of order 5(4)~\cite{dormand1980family} with different error tolerance settings.

We find that Euler-Heun achieves the best distributional results (KID and FID), Euler-Maruyama achieves the best perceptual result (LPIPS), and the best SSIM and PSNR values are reached by the ODE sampler with a tolerance of $10^{-3}$, though the differences are not large overall. The ODE sampler with tolerance $10^{-1}$ performs markedly worse than all other settings, which we found to be due to residual Gaussian noise being left in the reconstruction.

Concerning the \acf{nfe}, Euler-Maruyama serves as the baseline with a fixed number of 100. Euler-Heun requires twice as many \acp{nfe}, whereas the ODE samplers use only between 20 and 50 \acp{nfe} depending on the tolerance setting. We argue that, for a good balance between computational efficiency and reconstruction quality, the ODE sampler with a tolerance of $10^{-3}$ is the best choice as it needs only 26 \acp{nfe} and reaches comparable values to Euler-Maruyama. However, since we set our focus on perceptual quality where Euler-Maruyama performs best, we used this sampler for all other experiments in this manuscript.

\subsection{Limitations}
\label{sec:limitations}

\begin{figure}
    \centering
    \includegraphics[width=\linewidth]{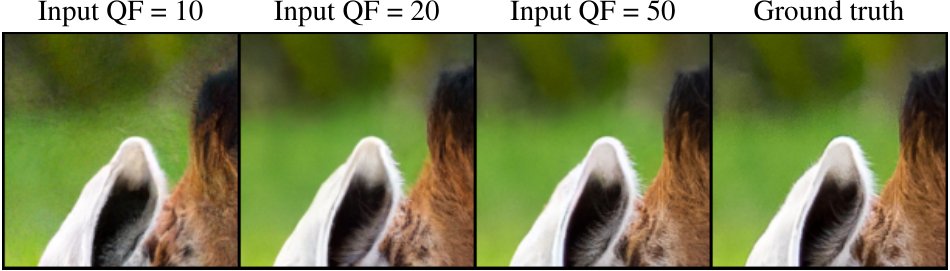}
    \caption{For very low \acp{qf} such as 10, DriftRec may erroneously generate 
    textured high-frequency content. This effect is however not present for \acp{qf} of 20 and higher.}
    \label{fig:d2kf2k-qf-comp}
\end{figure}

Our proposed method DriftRec has some limitations that we would like to point out. First, DriftRec implicitly assumes that the clean image $\x_0$ and the corrupted image $\y$ are reasonably similar to each other, so that performing per-pixel interpolation between them through the diffusion process -- see \cref{eq:mu-ouve,eq:mu-tsdve} -- is sensible. While single JPEG compression at medium-to-high \ac{qf} should generally fulfill this, extreme corruptions such as \acp{qf} less than 5 lead to large per-pixel distances between $\x_0$ and $\y$, and we found DriftRec to generate significant artifacts in these scenarios. However, we are not aware of other works that show convincing results for such low \acp{qf}, and such reconstruction tasks are also severely ill-posed due to extreme loss of information. We defer to concurrent work \cite{luo2023image} as well as our works in the speech domain \cite{richter2022journal,peer2023diffphase} to show that a variety of other restoration tasks can be solved with a similar approach and that this assumption is not a limiting factor in practice.

Second, we found that the reconstructed images by DriftRec sometimes exhibit slight global color shifts, e.g., which we hypothesize to be due to the additional signal power from the Gaussian noise added throughout the reverse sampling process. Our method has no explicit mechanism to control and suppress such shifts: slightly color-shifted images still have a high probability under the learned distribution. We thus applied a global color correction after sampling, by shifting each RGB channel in the reconstructed image to have the same mean as in the JPEG-compressed image. Doing so is reasonable as even heavy JPEG compression preserves global color information well, and this only relies on information in the corrupted image. We leave further improvements of this to future work.

Finally, we note that DriftRec may sometimes generate overly textured content for difficult input data. In \cref{fig:d2kf2k-qf-comp}, we show results for the same image with different input \acp{qf}. At \ac{qf} 10, the network generates fine structures where the background should be blurred and adds hair-like details not present in the ground-truth image. Note that for \ac{qf} 20 and 50, these generation artifacts are not present anymore. We expect that this issue can be effectively suppressed with larger networks, further research into noise schedules, or post-filtering techniques.

\subsection{Feature Maps}
\begin{figure}
    \centering
    \includegraphics[width=\linewidth]{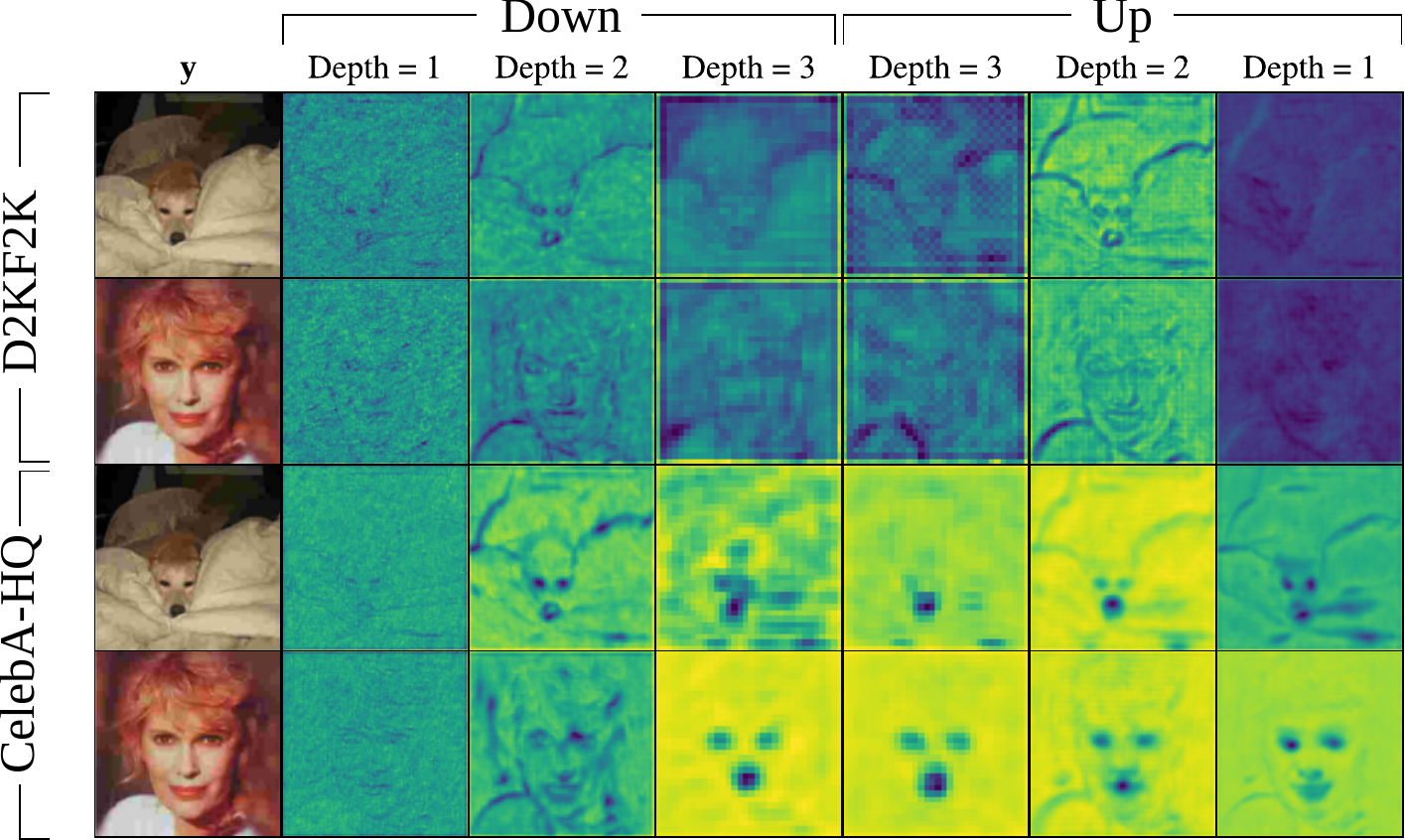}
    \caption{Activations (averaged over all channels) of the first convolution layer in selected ResNet blocks of the score network, evaluated at diffusion time $t=1$ for two DriftRec models trained on D2KF2K (top) and CelebA-HQ (bottom).
    }
    \label{fig:activations}
\end{figure}

In \cref{fig:activations}, we show feature maps generated from a forward pass through two DriftRec models, one trained on D2KF2K and the other on \celeba{}. We pass two example images at JPEG \ac{qf} 10 through both models at the diffusion process time $t=1$. For each listed depth of the NCSN++ U-Net, we show the average over all channels of the first convolution layer within the last ResNet block (see \cite{richter2022journal} for a visual overview of the architecture). One can see stark differences between the activations of the two models: The CelebA-HQ model produces feature maps that exhibit face-specific features, in particular eye and mouth positions, and even produces such feature maps for a non-human (dog) face. The D2KF2K model shows more generic features representing edge information. This indicates that our models learn and make use of dataset-specific high-level features, if available.

\subsection{Runtime performance and model size}
As our method is based on diffusion models, which use iterative methods for sampling, it also requires multiple calls to the same DNN to retrieve a single reconstruction. We used $N=100$ iterations in this work with the Euler-Maruyama sampler~\cite{kloeden2011numerical}, requiring the same number of $N_{\text{DNN}}=100$ DNN evaluations. On a \emph{NVIDIA GeForce RTX 2080 Ti} consumer GPU, we found this combination of sampler and DNN to require around 70 seconds to reconstruct a batch of sixteen $256\times{}256$ images, resulting in a total runtime of about 5.5 hours for all 4500 images in the \celeba{} test set. On the same hardware and dataset,  QGAC~\cite{ehrlich2020quantization} requires only 310 seconds in total, FBCNN~\cite{jiang2021towards} takes 279 seconds, and the regression baseline takes 400 seconds, with all three methods using a single DNN pass for each image.

However, since we derived our diffusion-based method within the continuous formalism of \acp{sde}, the choice of sampling procedure is highly flexible and does not require retraining of the DNN. Recent research on diffusion models has made significant progress on sampling speed~\cite{zhang2022fast,karras2022elucidating,salimans2022progressive}. These techniques can in principle be combined with our method but may require some careful re-derivation since works typically make assumptions about the \ac{sde} drift term $\mathbf f$ to be zero (\ac{ve} \ac{sde}) or a simple multiplicative scaling of $\x_t$ (\ac{vp} \ac{sde}), which does not hold for our \acp{sde}.

Finally, pointing to \cref{tab:model-size}, we note that DriftRec has the lowest number of DNN parameters compared to the other methods, and virtually the same number of DNN parameters as our regression baseline since it uses the same DNN architecture with very minor modifications, see \cref{sec:architecture-and-training}.

\section{Conclusions}
Based on our work from the speech processing literature~\cite{welker2022speech,richter2022journal}, we propose \emph{DriftRec}, a method based on an elegant change to \ac{sde}-based diffusion models that adapts them to image restoration tasks. In contrast to existing diffusion-based restoration methods~\cite{kawar2022denoising,kawar2022jpeg,daras2023soft,bansal2022cold}, DriftRec is applicable to corruption operators that are both nonlinear and unavailable in closed form during reconstruction, only requiring a dataset of paired images during training and being able to perform blind restoration, as demonstrated for a variety of image reconstruction tasks in recent concurrent work~\cite{luo2023image}. We utilize these properties of DriftRec to attack blind JPEG restoration for low quality factors (\ac{qf}~$\leq 30$), though our method is trained for and applicable to all quality factors 0--100. We compare three variants of DriftRec against an $L_2$ regression baseline using the same DNN architecture as well as recent state-of-the-art methods and find that DriftRec restores images of significantly higher perceptual quality and models the ground-truth image distribution significantly more faithfully. We find that DriftRec has a useful ability to generalize to more complex JPEG restoration problems: unaligned double JPEG compression, and restoration of JPEG images found online where the exact quality factors and further processing are unknown. We show furthermore that DriftRec can improve downstream tasks such as image classification. We explore the behavior of different samplers to achieve reconstructions with higher efficiency, and show that averaging multiple samples can be employed to trade per-pixel distortion for perceptual quality.

While DriftRec currently uses between 20 and 100 times as many DNN passes than the compared methods due to the iterative nature of sampling with diffusion models, we expect that recent progress on sampling efficiency for diffusion models, e.g.~\cite{watson2021learning,lemercier2023storm,lay2023reducing,song2023consistency}, will allow this number to drastically decrease and make the approach competitive in terms of runtime. Our idea of adapting the diffusion process can in principle be combined with most other techniques for designing, improving, or specializing diffusion models, and therefore constitutes another building block in the growing diffusion model toolbox.

\section*{Acknowledgments}
We acknowledge the support by DASHH (Data Science in Hamburg - HELMHOLTZ Graduate School for the Structure of Matter) with the Grant-No. HIDSS-0002. This research was supported in part through the Maxwell computational resources operated at Deutsches Elektronen-Synchrotron DESY, Hamburg, Germany.

{
\bibliographystyle{IEEEtran}
\bibliography{IEEEabrv,bib}

\begin{thebibliography}{10}
\providecommand{\url}[1]{#1}
\csname url@samestyle\endcsname
\providecommand{\newblock}{\relax}
\providecommand{\bibinfo}[2]{#2}
\providecommand{\BIBentrySTDinterwordspacing}{\spaceskip=0pt\relax}
\providecommand{\BIBentryALTinterwordstretchfactor}{4}
\providecommand{\BIBentryALTinterwordspacing}{\spaceskip=\fontdimen2\font plus
\BIBentryALTinterwordstretchfactor\fontdimen3\font minus
  \fontdimen4\font\relax}
\providecommand{\BIBforeignlanguage}[2]{{%
\expandafter\ifx\csname l@#1\endcsname\relax
\typeout{** WARNING: IEEEtran.bst: No hyphenation pattern has been}%
\typeout{** loaded for the language `#1'. Using the pattern for}%
\typeout{** the default language instead.}%
\else
\language=\csname l@#1\endcsname
\fi
#2}}
\providecommand{\BIBdecl}{\relax}
\BIBdecl

\bibitem{dhariwal2021diffusion}
P.~Dhariwal and A.~Nichol, ``Diffusion models beat {GANs} on image synthesis,''
  \emph{Advances in Neural Inf. Proc. Systems (NeurIPS)}, vol.~34, pp.
  8780--8794, 2021.

\bibitem{ramesh2022hierarchical}
A.~Ramesh, P.~Dhariwal, A.~Nichol, C.~Chu, and M.~Chen, ``Hierarchical
  text-conditional image generation with {CLIP} latents,'' \emph{arXiv preprint
  arXiv:2204.06125}, 2022.

\bibitem{rombach2022high}
R.~Rombach, A.~Blattmann, D.~Lorenz, P.~Esser, and B.~Ommer, ``High-resolution
  image synthesis with latent diffusion models,'' in \emph{IEEE/CVF Conf. on
  Computer Vision and Pattern Recognition (CVPR)}, June 2022, pp.
  10\,684--10\,695.

\bibitem{kong2021diffwave}
Z.~Kong, W.~Ping, J.~Huang, K.~Zhao, and B.~Catanzaro, ``{DiffWave}: A
  versatile diffusion model for audio synthesis,'' \emph{Int. Conf. on Learning
  Representations (ICLR)}, pp. 1--12, 2021.

\bibitem{richter2022journal}
J.~Richter, S.~Welker, J.-M. Lemercier, B.~Lay, and T.~Gerkmann, ``Speech
  enhancement and dereverberation with diffusion-based generative models,''
  \emph{IEEE Trans. on Audio, Speech, and Language Proc. (TASLP)}, p.
  2351–2364, 2023.

\bibitem{huang2022fastdiff}
R.~Huang, M.~W.~Y. Lam, J.~Wang, D.~Su, D.~Yu, Y.~Ren, and Z.~Zhao,
  ``{FastDiff}: A fast conditional diffusion model for high-quality speech
  synthesis,'' in \emph{IJCAI}, 2022, pp. 4157--4163.

\bibitem{sheffer2022hear}
R.~Sheffer and Y.~Adi, ``I hear your true colors: Image guided audio
  generation,'' in \emph{IEEE Int. Conf. on Acoustics, Speech and Signal Proc.
  (ICASSP)}.\hskip 1em plus 0.5em minus 0.4em\relax IEEE, 2023, pp. 1--5.

\bibitem{bansal2022cold}
A.~Bansal, E.~Borgnia, H.-M. Chu, J.~S. Li, H.~Kazemi, F.~Huang, M.~Goldblum,
  J.~Geiping, and T.~Goldstein, ``Cold diffusion: Inverting arbitrary image
  transforms without noise,'' \emph{arXiv preprint arXiv:2208.09392}, 2022.

\bibitem{daras2023soft}
G.~Daras, M.~Delbracio, H.~Talebi, A.~Dimakis, and P.~Milanfar, ``Soft
  diffusion: Score matching with general corruptions,'' \emph{Transactions on
  Machine Learning Research}, 2023.

\bibitem{kawar2022denoising}
B.~Kawar, M.~Elad, S.~Ermon, and J.~Song, ``Denoising diffusion restoration
  models,'' in \emph{NeurIPS}, 2022, pp. 23\,593--23\,606.

\bibitem{song2021sde}
Y.~Song, J.~Sohl-Dickstein, D.~P. Kingma, A.~Kumar, S.~Ermon, and B.~Poole,
  ``Score-based generative modeling through stochastic differential
  equations,'' \emph{Int. Conf. on Learning Representations (ICLR)}, pp. 1--12,
  2021.

\bibitem{welker2022speech}
S.~Welker, J.~Richter, and T.~Gerkmann, ``Speech enhancement with score-based
  generative models in the complex {STFT} domain,'' \emph{ISCA Interspeech},
  pp. 2928--2932, 2022.

\bibitem{welker2022blind}
S.~Welker, H.~N. Chapman, and T.~Gerkmann, ``Blind drifting: Diffusion models
  with a linear {SDE} drift term for blind image restoration tasks,'' in
  \emph{NeurIPS 2022 Workshop: The Symbiosis of Deep Learning and Differential
  Equations II (Spotlight)}, 2022, pp. 1--5.

\bibitem{kawar2022jpeg}
B.~Kawar, J.~Song, S.~Ermon, and M.~Elad, ``{JPEG} artifact correction using
  {Denoising Diffusion Restoration Models},'' in \emph{NeurIPS 2022 Workshop on
  Score-Based Methods}, 2022, pp. 1--7.

\bibitem{saharia2022palette}
C.~Saharia, W.~Chan, H.~Chang, C.~Lee, J.~Ho, T.~Salimans, D.~Fleet, and
  M.~Norouzi, ``Palette: Image-to-image diffusion models,'' in \emph{ACM
  SIGGRAPH 2022 Conference Proceedings}, 2022, pp. 1--10.

\bibitem{jiang2021towards}
J.~Jiang, K.~Zhang, and R.~Timofte, ``Towards flexible blind {JPEG} artifacts
  removal,'' in \emph{ICCV}, 2021, pp. 4997--5006.

\bibitem{ehrlich2020quantization}
M.~Ehrlich, L.~Davis, S.-N. Lim, and A.~Shrivastava, ``Quantization guided
  {JPEG} artifact correction,'' in \emph{ECCV}.\hskip 1em plus 0.5em minus
  0.4em\relax Springer, 2020, pp. 293--309.

\bibitem{fu2021model}
X.~Fu, M.~Wang, X.~Cao, X.~Ding, and Z.-J. Zha, ``A model-driven deep unfolding
  method for {JPEG} artifacts removal,'' \emph{IEEE Transactions on Neural
  Networks and Learning Systems}, vol.~33, no.~11, pp. 6802--6816, 2021.

\bibitem{fu2021learning}
X.~Fu, X.~Wang, A.~Liu, J.~Han, and Z.-J. Zha, ``Learning dual priors for
  {JPEG} compression artifacts removal,'' in \emph{Proceedings of the IEEE/CVF
  International Conference on Computer Vision}, 2021, pp. 4086--4095.

\bibitem{luo2023image}
Z.~Luo, F.~K. Gustafsson, Z.~Zhao, J.~Sj{\"o}lund, and T.~B. Sch{\"o}n, ``Image
  restoration with mean-reverting stochastic differential equations,'' in
  \emph{Int. Conf. on Machine Learning (ICML)}, 2023, pp. 23\,045--23\,066.

\bibitem{welker2022driftrec}
S.~Welker, H.~N. Chapman, and T.~Gerkmann, ``Driftrec: Adapting diffusion
  models to blind image restoration tasks,'' \emph{arXiv preprint
  arXiv:2211.06757}, 2022.

\bibitem{liu2023i2sb}
G.-H. Liu, A.~Vahdat, D.-A. Huang, E.~Theodorou, W.~Nie, and A.~Anandkumar,
  ``{I$^2$SB}: {Image-to-Image Schr{\"o}dinger Bridge},'' \emph{Int. Conf. on
  Learning Representations (ICLR)}, pp. 22\,042--22\,062, 2023.

\bibitem{delbracio2023inversion}
\BIBentryALTinterwordspacing
M.~Delbracio and P.~Milanfar, ``Inversion by direct iteration: An alternative
  to denoising diffusion for image restoration,'' \emph{Transactions on Machine
  Learning Research}, 2023, featured Certification. [Online]. Available:
  \url{https://openreview.net/forum?id=VmyFF5lL3F}
\BIBentrySTDinterwordspacing

\bibitem{anderson1982reverse}
B.~D.~O. Anderson, ``Reverse-time diffusion equation models,'' \emph{Stochastic
  Processes and their Applications}, vol.~12, no.~3, pp. 313--326, 1982.

\bibitem{sarkka2019sde}
S.~Särkkä and A.~Solin, \emph{Applied Stochastic Differential
  Equations}.\hskip 1em plus 0.5em minus 0.4em\relax {Cambridge University
  Press}, 2019.

\bibitem{kloeden2011numerical}
P.~Kloeden and E.~Platen, \emph{Numerical Solution of Stochastic Differential
  Equations}, ser. Stochastic Modelling and Applied Probability.\hskip 1em plus
  0.5em minus 0.4em\relax Springer Berlin Heidelberg, 2011.

\bibitem{meng2022sdedit}
C.~Meng, Y.~He, Y.~Song, J.~Song, J.~Wu, J.-Y. Zhu, and S.~Ermon, ``{SDE}dit:
  Guided image synthesis and editing with stochastic differential equations,''
  in \emph{Int. Conf. on Learning Representations (ICLR)}, 2022, pp. 1--14.

\bibitem{song2019generative}
Y.~Song and S.~Ermon, ``Generative modeling by estimating gradients of the data
  distribution,'' \emph{Advances in Neural Inf. Proc. Systems (NeurIPS)},
  vol.~32, pp. 1--13, 2019.

\bibitem{berner2022optimal}
J.~Berner, L.~Richter, and K.~Ullrich, ``An optimal control perspective on
  diffusion-based generative modeling,'' in \emph{NeurIPS 2022 Workshop on
  Score-Based Methods}, 2022, pp. 1--6.

\bibitem{vincent2011connection}
P.~Vincent, ``A connection between score matching and denoising autoencoders,''
  \emph{Neural Computation}, vol.~23, no.~7, pp. 1661--1674, 2011.

\bibitem{karras2018progressive}
T.~Karras, T.~Aila, S.~Laine, and J.~Lehtinen, ``Progressive growing of {GAN}s
  for improved quality, stability, and variation,'' in \emph{Int. Conf. on
  Learning Representations (ICLR)}, 2018, pp. 1--12.

\bibitem{agustsson2017div2k}
E.~Agustsson and R.~Timofte, ``{NTIRE 2017 Challenge} on single image
  super-resolution: Dataset and study,'' in \emph{CVPRW}, July 2017.

\bibitem{timofte2017flickr2k}
R.~Timofte, E.~Agustsson, L.~Van~Gool, M.-H. Yang, L.~Zhang, B.~Lim
  \emph{et~al.}, ``{NTIRE 2017 Challenge} on single image super-resolution:
  Methods and results,'' in \emph{CVPRW}, July 2017.

\bibitem{sheikh2006statistical}
H.~R. Sheikh, M.~F. Sabir, and A.~C. Bovik, ``A statistical evaluation of
  recent full reference image quality assessment algorithms,'' \emph{IEEE
  Transactions on image processing}, vol.~15, no.~11, pp. 3440--3451, 2006.

\bibitem{clark2015pillow}
\BIBentryALTinterwordspacing
A.~Clark, ``Pillow ({PIL} fork) documentation,'' 2015. [Online]. Available:
  \url{https://buildmedia.readthedocs.org/media/pdf/pillow/latest/pillow.pdf}
\BIBentrySTDinterwordspacing

\bibitem{loshchilov2018decoupled}
I.~Loshchilov and F.~Hutter, ``Decoupled weight decay regularization,'' in
  \emph{Int. Conf. on Learning Representations (ICLR)}, 2019, pp. 1--10.

\bibitem{binkowski2018demystifying}
M.~Bińkowski, D.~J. Sutherland, M.~Arbel, and A.~Gretton, ``Demystifying {MMD}
  {GAN}s,'' in \emph{Int. Conf. on Learning Representations (ICLR)}, 2018, pp.
  1--15.

\bibitem{FIDref}
M.~Heusel, H.~Ramsauer, T.~Unterthiner, B.~Nessler, and S.~Hochreiter, ``{GAN}s
  trained by a two time-scale update rule converge to a local {Nash}
  equilibrium,'' in \emph{Advances in Neural Inf. Proc. Systems (NeurIPS)},
  vol.~30, 2017, pp. 1--12.

\bibitem{wang2004image}
Z.~Wang, A.~C. Bovik, H.~R. Sheikh, and E.~P. Simoncelli, ``Image quality
  assessment: from error visibility to structural similarity,'' \emph{IEEE
  Trans. on Image Proc.}, vol.~13, no.~4, pp. 600--612, 2004.

\bibitem{zhang2018perceptual}
R.~Zhang, P.~Isola, A.~A. Efros, E.~Shechtman, and O.~Wang, ``The unreasonable
  effectiveness of deep features as a perceptual metric,'' in \emph{IEEE/CVF
  Conf. on Computer Vision and Pattern Recognition (CVPR)}, 2018, pp. 586--595.

\bibitem{van2014scikit}
S.~Van~der Walt, J.~L. Sch{\"o}nberger, J.~Nunez-Iglesias, F.~Boulogne, J.~D.
  Warner, N.~Yager, E.~Gouillart, and T.~Yu, ``scikit-image: image processing
  in {Python},'' \emph{PeerJ}, vol.~2, p. e453, 2014.

\bibitem{kastryulin2022piq}
S.~Kastryulin, J.~Zakirov, D.~Prokopenko, and D.~V. Dylov, ``{PyTorch} image
  quality: Metrics for image quality assessment,'' \emph{arXiv preprint
  arXiv:2208.14818}, 2022.

\bibitem{dahl2017pixel}
R.~Dahl, M.~Norouzi, and J.~Shlens, ``Pixel recursive super resolution,'' in
  \emph{ICCV}, 2017, pp. 5439--5448.

\bibitem{ledig2017photo}
C.~Ledig, L.~Theis, F.~Husz{\'a}r, J.~Caballero, A.~Cunningham, A.~Acosta,
  A.~Aitken, A.~Tejani, J.~Totz, Z.~Wang \emph{et~al.}, ``Photo-realistic
  single image super-resolution using a generative adversarial network,'' in
  \emph{IEEE/CVF Conf. on Computer Vision and Pattern Recognition (CVPR)},
  2017, pp. 4681--4690.

\bibitem{mier2021deep}
J.~C. Mier, E.~Huang, H.~Talebi, F.~Yang, and P.~Milanfar, ``Deep perceptual
  image quality assessment for compression,'' in \emph{ICIP}.\hskip 1em plus
  0.5em minus 0.4em\relax IEEE, 2021, pp. 1484--1488.

\bibitem{blau2018perception}
Y.~Blau and T.~Michaeli, ``The perception-distortion tradeoff,'' in
  \emph{IEEE/CVF Conf. on Computer Vision and Pattern Recognition (CVPR)}, June
  2018.

\bibitem{tadala2012psnrb}
T.~Tadala and S.~E.~V. Narayana, ``A novel {PSNR-B} approach for evaluating the
  quality of deblocked images,'' \emph{IOSR Journal of Computer Engineering},
  vol.~4, no.~5, pp. 40--49, 2012.

\bibitem{roberts2012modify}
A.~Roberts, ``Modify the improved {Euler} scheme to integrate stochastic
  differential equations,'' \emph{arXiv preprint arXiv:1210.0933}, 2012.

\bibitem{dormand1980family}
J.~R. Dormand and P.~J. Prince, ``A family of embedded runge-kutta formulae,''
  \emph{Journal of computational and applied mathematics}, vol.~6, no.~1, pp.
  19--26, 1980.

\bibitem{peer2023diffphase}
T.~Peer, S.~Welker, and T.~Gerkmann, ``Diffphase: Generative diffusion-based
  {STFT} phase retrieval,'' in \emph{ICASSP 2023-2023 IEEE International
  Conference on Acoustics, Speech and Signal Processing (ICASSP)}.\hskip 1em
  plus 0.5em minus 0.4em\relax IEEE, 2023, pp. 1--5.

\bibitem{zhang2022fast}
Q.~Zhang and Y.~Chen, ``Fast sampling of diffusion models with exponential
  integrator,'' in \emph{NeurIPS 2022 Workshop on Score-Based Methods}, 2022,
  pp. 1--6.

\bibitem{karras2022elucidating}
T.~Karras, M.~Aittala, T.~Aila, and S.~Laine, ``Elucidating the design space of
  diffusion-based generative models,'' in \emph{Advances in Neural Inf. Proc.
  Systems (NeurIPS)}, 2022, pp. 26\,565--26\,577.

\bibitem{salimans2022progressive}
T.~Salimans and J.~Ho, ``Progressive distillation for fast sampling of
  diffusion models,'' in \emph{Int. Conf. on Learning Representations (ICLR)},
  2022, pp. 1--12.

\bibitem{watson2021learning}
D.~Watson, W.~Chan, J.~Ho, and M.~Norouzi, ``Learning fast samplers for
  diffusion models by differentiating through sample quality,'' in \emph{Int.
  Conf. on Learning Representations (ICLR)}, 2021, pp. 1--12.

\bibitem{lemercier2023storm}
J.-M. Lemercier, J.~Richter, S.~Welker, and T.~Gerkmann, ``{StoRM}: A
  diffusion-based stochastic regeneration model for speech enhancement and
  dereverberation,'' \emph{IEEE Trans. on Audio, Speech, and Language Proc.
  (TASLP)}, pp. 2724--2737, 2023.

\bibitem{lay2023reducing}
B.~Lay, S.~Welker, J.~Richter, and T.~Gerkmann, ``Reducing the prior mismatch
  of stochastic differential equations for diffusion-based speech
  enhancement,'' \emph{ISCA Interspeech}, pp. 3809--3813, 2023.

\bibitem{song2023consistency}
Y.~Song, P.~Dhariwal, M.~Chen, and I.~Sutskever, ``Consistency models,''
  \emph{Int. Conf. on Machine Learning (ICML)}, pp. 32\,211--32\,252, 2023.

\end{thebibliography}
}

\newpage

\begin{IEEEbiography}[{\includegraphics[width=1in,height=1.25in,clip,keepaspectratio]{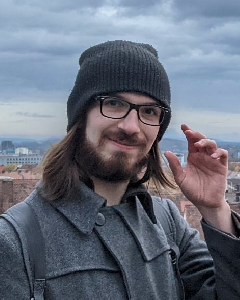}}]{Simon Welker} (\href{mailto:simon.welker@uni-hamburg.de}{simon.welker@uni-hamburg.de})
received a B.Sc. in Computing in Science (2019) and M.Sc. in Bioinformatics (2021) from Universität Hamburg, Germany. He is currently pursuing a doctoral degree in the groups of Prof. Timo Gerkmann (Signal Processing, Universität Hamburg) and Prof. Henry N. Chapman (Center for Free-Electron Laser Science, DESY, Hamburg), researching machine learning techniques for solving inverse problems that arise in speech processing and X-ray imaging.
\end{IEEEbiography}

\begin{IEEEbiography}[{\includegraphics[width=1in,height=1.25in,clip,keepaspectratio]{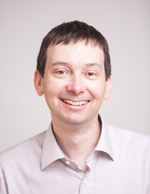}}]{Henry N. Chapman} (\href{mailto:henry.chapman@desy.de}{henry.chapman@desy.de}) is a director of the Center for Free-Electron Laser Science at the Deutsches Elektronen-Synchrotron and the University of Hamburg in Germany. He earned his Ph.D. degree in X-ray optics at the University of Melbourne, Australia. He has developed methods in coherent X-ray imaging, which began at Stony Brook University in New York. At Lawrence Livermore National Laboratory in California, he led a team to demonstrate that an intense X-ray free-electron laser pulse could outrun radiation damage. He continued this work to the atomic scale with the method of serial femtosecond X-ray crystallography, which promises to overcome current bottlenecks in protein structure determination. His current research is focused on developing this method and extending it to the smallest possible crystals, i.e., single molecules.
\end{IEEEbiography}

\begin{IEEEbiography}[{\includegraphics[width=1in,height=1.25in,clip,keepaspectratio]{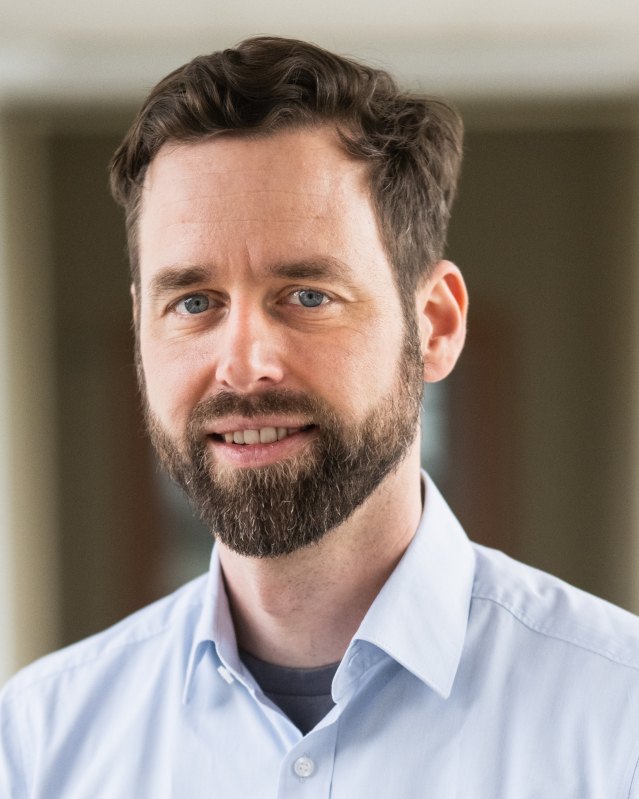}}]{Timo Gerkmann} (\href{mailto:timo.gerkmann@uni-hamburg.de}{timo.gerkmann@uni-hamburg.de})
(S’08–M’10–SM’15) is a professor for Signal Processing at the Universität Hamburg, Germany. He has previously held positions at Technicolor Research \& Innovation in Germany, the University of Oldenburg in Germany, KTH Royal Institute of Technology in Sweden, Ruhr-Universität Bochum in Germany, and Siemens Corporate Research in Princeton, NJ, USA. His main research interests are on statistical signal processing and machine learning for speech and audio applied to communication devices, hearing instruments, audio-visual media, and human-machine interfaces. Timo Gerkmann served as a member of the IEEE Signal Processing Society Technical Committee on Audio and Acoustic Signal Processing (2018-2023), as an Associate Editor (2019-2022) and since 2022 serves as a Senior Area Editor of the IEEE/ACM Transactions on Audio, Speech and Language Processing. He received the VDE ITG award 2022.
\end{IEEEbiography}

\vfill

\end{document}